\documentclass[journal]{IEEEtran}

\usepackage[utf8]{inputenc}
\usepackage[T1]{fontenc}
\usepackage[scaled=0.92]{helvet}  
\usepackage[noadjust]{cite}

\usepackage{graphicx}
\usepackage{array}
\newcolumntype{M}[1]{>{\centering\arraybackslash}m{#1}}
\usepackage{tabularx}
\usepackage{dblfloatfix}
   
\usepackage{amsmath}
\usepackage{newtxtext, newtxmath}
\usepackage{dsfont}
\usepackage{hhline,booktabs}
\usepackage{tablefootnote}
\usepackage{enumitem}
\usepackage[T1]{fontenc}
\usepackage{adjustbox}
\usepackage{booktabs, multirow}
\usepackage{amsmath}

\usepackage[scaled=0.9]{DejaVuSansMono}
\usepackage{listings}
\usepackage{amsfonts}

\usepackage{amsthm,thmtools,xcolor}
\declaretheoremstyle[
  headfont=\color{red}\normalfont\bfseries,
  bodyfont=\color{red}\normalfont\itshape,
]{colored}

\usepackage{bm}
\usepackage{url}
\usepackage{mathtools}
\usepackage{balance}
\usepackage{subcaption}
\allowdisplaybreaks
\usepackage{pgfplotstable}

\usepackage{tkz-graph}
\tikzset{
  LabelStyle/.style = { rectangle, rounded corners, draw=black,dashed,
                        minimum width = 2em, fill=red!20,
                        text = black, font = \bfseries },
  VertexStyle/.append style = {ultra thick,draw=blue!75,fill=blue!20,shape=circle, inner sep=2pt,
                                font = \Large\bfseries},
  EdgeStyle/.append style = {ultra thick,draw=red!70,>={Stealth[black]}} }

\usepackage{algorithm}
\usepackage[noend]{algpseudocode}
\usepackage{etoolbox}
\usepackage[mathscr]{euscript}
\usepackage{calc}
\usepackage{pgfplots}
\usepackage{tikz}
\usepackage{setspace}
\usepackage{pgfplotstable}

\usepackage[font=small]{caption}
\DeclareMathAlphabet{\pazocal}{OMS}{zplm}{m}{n}
\DeclareSymbolFont{missing}{OML}{cmr}{m}{n}
\DeclareMathSymbol{\ell}{\mathord}{missing}{'140}

\usetikzlibrary{calc}
\usetikzlibrary{patterns}
\usetikzlibrary{spy,backgrounds}
\pgfplotsset{grid style={dotted,gray}}

\newcommand{\maximize}{%
  \mathopen{}\operatorname*{maximize}%
}

\newcommand{\subjto}{\textup{subject to}}

\newcounter{problem}
\newcounter{save@equation}
\newcounter{save@problem}

\newtheorem{lemma}{Lemma}
\newtheorem{theorem}{Theorem}

\newtheorem{definition}{Definition}

\newtheorem{remark}{Remark}
\newlength{\depthofsumsign}
\newlength{\totalheightofsumsign}
\newlength{\heightanddepthofargument}

\makeatletter

\tikzset{reset label anchor/.code={%
    \let\tikz@auto@anchor=\pgfutil@empty
    \def\tikz@anchor{#1}
  },
  reset label anchor/.default=center
}

\let\save@mathaccent\mathaccent
\newcommand*\if@single[3]{%
  \setbox0\hbox{${\mathaccent"0362{#1}}^H$}%
  \setbox2\hbox{${\mathaccent"0362{\kern0pt#1}}^H$}%
  \ifdim\ht0=\ht2 #3\else #2\fi
}

\newenvironment{problem}
{\setcounter{problem}{\value{save@problem}}%
  \setcounter{save@equation}{\value{equation}}%
  \let\c@equation\c@problem
  \renewcommand{\theequation}{P\arabic{equation}}%
  \subequations
}
{\endsubequations
  \setcounter{save@problem}{\value{equation}}%
  \setcounter{equation}{\value{save@equation}}%
}

\algnewcommand{\LineComment}[1]{\Statex \hskip\ALG@thistlm
  \(\triangleright\) #1}
\def\BState{\State\hskip-\ALG@thistlm}

\algdef{SE}[DOWHILE]{Do}{EndDoWhile}{\algorithmicdo}[1]{\algorithmicwhile\ #1}%

   \tikzset{nomorepostaction/.code=\let\tikz@postactions\pgfutil@empty}

   \long\def\ifnodedefined#1#2#3{%
   \@ifundefined{pgf@sh@ns@#1}{#3}{#2}%
 }
\usetikzlibrary{decorations.pathmorphing,arrows.meta,positioning}
\tikzstyle{printersafe}=[decoration={amplitude=0pt}]

\pgfplotsset{
  every axis plot/.append style={line width=1pt},
}

\def\ps@IEEEtitlepagestyle{
  \def\@oddfoot{\mycopyrightnotice}
  \def\@evenfoot{}
}
\def\mycopyrightnotice{
  {\footnotesize
  \begin{minipage}{\textwidth}
  \centering
Copyright (c) 2021 IEEE. Personal use of this material is permitted. However, permission to use this material for any other purposes must be obtained from the IEEE by sending a request to pubs-permissions@ieee.org.
  \end{minipage}
  }
}

\makeatother

\usetikzlibrary{automata}
\usetikzlibrary{calc,fit,shapes.geometric}
\pgfdeclarelayer{signal}
\pgfsetlayers{signal,main}
\tikzstyle{printersafe}=[segment amplitude=0 pt]
\usetikzlibrary{arrows}

\usetikzlibrary{calc}
\usetikzlibrary{decorations.pathreplacing,decorations.markings,shapes.geometric}
\tikzset{naming/.style={align=center}}
\tikzset{
  every pin/.style={rectangle,rounded corners=3pt,font=\footnotesize},
  small dot/.style={fill=black,circle,scale=0.5}
}

\tikzset{
  invisible/.style={opacity=0},
  visible on/.style={alt={#1{}{invisible}}},
  alt/.code args={<#1>#2#3}{%
    \alt<#1>{\pgfkeysalso{#2}}{\pgfkeysalso{#3}} 
  },
}

\colorlet{sky blue}{blue!60!cyan!75!black}
\colorlet{dark blue}{blue!50!cyan}
\colorlet{chameleon}{olive!75!green}
\tikzset{signal/.style={->,draw=black, line width=0.05em, dashed,printersafe}}

\newcommand*\upper[1]{\bar{#1}}
\newsavebox{\mybox}
\pgfplotsset{compat=1.16}
\begin{document}
\bstctlcite{M2Mv1}

\title{Massive IoT Access with NOMA in 5G Networks and Beyond using Online Competitiveness and Learning}

\author{\IEEEauthorblockN{Zoubeir~Mlika,~\IEEEmembership{Member,~IEEE} and Soumaya~Cherkaoui,~\IEEEmembership{Senior~Member,~IEEE}} \\ \IEEEauthorblockA{Department of Electrical and Computer Engineering, University of Sherbrooke\\ zoubeir.mlika@usherbrooke.ca, soumaya.cherkaoui@usherbrooke.ca}}%

\maketitle

\begin{abstract}
  This paper studies the problem of online user grouping, scheduling, and power allocation for massive Internet of things (IoT) access in beyond 5G networks using non-orthogonal multiple access (NOMA). NOMA has been identified as a promising technology to accommodate a large number of devices using a limited number of radio resources. In this work, the objective is to maximize the number of served devices while allocating their transmission powers such that their real-time requirements as well as their limited operating energy are respected. First, we formulate the problem as a mixed-integer non-linear program (MINLP) that can be transformed to MILP for some special cases. Second, we study its NP-hardness in different cases. Then, by dividing the problem into multiple NOMA grouping and scheduling subproblems, an efficient online competitive algorithm is proposed to solve each subproblem. Next, we show how to use the proposed online algorithm as a black box and how to combine the obtained solutions to each subproblem in a reinforcement learning setting to obtain the power allocation for each NOMA group. Our analyses are supplemented by simulation results to illustrate the performance of the proposed algorithms in comparison to optimal and state-of-the-art methods.
\end{abstract}

\begin{IEEEkeywords}
  Internet of Things, Machine-to-Machine, NOMA, Power Allocation, NP-hardness, Competitive Algorithms, Reinforcement Learning.
\end{IEEEkeywords}

\newcommand{\describeContent}[1]{%
\begingroup%
\let\thefootnote\relax%
\footnotetext{#1}%
\endgroup%
}

\IEEEpeerreviewmaketitle

\section{Introduction}\label{section:introduction}
Tens of billions of objects will be connected to the Internet soon. These objects form the well-known Internet of things (IoT)~\cite{7835337,ALOQAILY2020254}, which is one of the promising applications in future wireless networks~\cite{zig,78353371,7132717}, including fifth-generation (5G) standard and beyond 5G (B5G). To realize IoT, machine-to-machine (M2M) communication is proposed where objects communicate with each other without (or with little) human interactions. The applications of M2M in IoT networks include smart cities, smart grids, industrial automation, health-care, intelligent transportation systems, to name only a few~\cite{8253977}. Due to the maturity of cellular networks and their ability to provide wide-area coverage, the integration of M2M communication with cellular networks can be viewed as a viable solution to the realization of such applications. For example, narrow-band IoT (NB-IoT)~\cite{7876968} is proposed by the 3rd generation partnership project as a cellular-based M2M communication network using the long term evolution standard.

The main requirement of cellular-based M2M communication is massive connectivity, i.e, a massive number of objects (or interchangeably called devices) communicating with each other through cellular networks need to be supported in the future~\cite{7736615,7165674}. For example, in~\cite{6477831}, it was expected that more than two billion objects will be directly connected to cellular networks. Besides massive connectivity, M2M traffic is generally different from traditional cellular traffic. It is sporadic and characterized by small-sized packets. Thus, maximizing the sum-rate is not the priority anymore in cellular-based M2M networks. Further, the cellular-based M2M networks have more stringent latency and energy-efficiency requirements. Consequently, rethinking the resource allocation methods is of great importance in such networks.

In this paper, we study an online resource allocation problem in cellular-based M2M networks. While most previous works study the sum-rate-related objectives and use stochastic Lyapunov framework to solve the problem, in this paper, we consider a different objective that is better suited to the massive IoT access problem in such networks. The objective is to maximize the number of served devices (NSD) or the number of successfully received packets while preserving their energy and satisfying their real-time requirements. To accommodate a large number of devices, the non-orthogonal multiple access (NOMA) technique is used. Note that it is well-known that maximizing the sum-rate can be achieved while serving only a few devices. This shows that both objectives might lead to very different problems and thus it illustrates the importance of optimizing the NSD in IoT networks. The problem is therefore to maximize the NSD while (i) grouping IoT devices into the available resource blocks (ii) scheduling IoT device transmission to respect their real-time requirements, and (iii) allocating IoT device transmission power to respect their limited operating energy levels. We solve the problem of device grouping and power allocation (GPA) using online computation~\cite{Borodin1998} and learning~\cite{MAL018} frameworks.

\subsection{Related Work}
Most previous works considered the problem of resource allocation in M2M networks from the perspective of either maximizing the sum-rate or minimizing the sum-power or other related objectives. Further, few research papers focus on the online competitive analysis and learning frameworks.

In~\cite{8663999}, the authors give a short survey for the problem of uplink grant in M2M networks. Comparison between coordinated and uncoordinated access is shown. The problem is formulated as a prediction problem: to reduce the delay, the devices do not have to send random access requests to the BS. Instead, the BS allocates its resources to the devices by predicting which device has packets to send. Further, the authors developed a two-stage solution based on machine learning. In~\cite{8644350}, a multi-armed bandit approach is proposed to solve the problem of fast uplink grant access in M2M networks. The objective is to maximize a utility function that is a combination of the data rate, the access delay, and the value of data packets. Since the set of possible actions is not known in advance, a sleeping multi-armed bandit technique is used. In~\cite{8370650}, the authors study the resource management problem in green IoT multi-hop networks. The IoT devices obtain energy through either grid power or energy harvesting. The problem is formulated as a stochastic optimization problem where the objective is to maximize a combination of a time-average network utility and an energy purchasing cost. Lyapunov optimization techniques are used to obtain a stable solution in large- and small-time scales. In~\cite{8318569}, the authors formulate a dynamic scheduling and power allocation problem in IoT networks using the NOMA technique. A stochastic optimization problem is formulated. The objective is to minimize the long-term average power consumption subject to the maximum transmission power, the scheduling, and the long average rate constraints. Well-known techniques are used to derive the Lyapunov function and the upper bound on the drift plus penalty. Then, the problem is transformed into a set of static optimization problems that are solved iteratively. Next, the branch and bound technique is used to solve the problem. In~\cite{7557079}, the authors study clustering and power allocation in NOMA systems to answer the following question: how to group the users into the resource blocks and how to allocate the transmission power to maximize the system throughput while guaranteeing the minimum rate requirements of the users and without exceeding the maximum transmission power. The solutions are divided into user clustering by developing algorithms that satisfy the successive interference cancellation (SIC) constraints, and power allocation using the Lagrangian method. In~\cite{8777301}, dynamic power control and user pairing problems are solved in delay-constrained multiple access networks with hybrid OMA and NOMA techniques. The objective is to minimize the long-term time-average transmission power while guaranteeing the stability of the queues and the minimum time-average rate. By fixing the user pairing, the power allocation is derived. Further, for a given power allocation, the user pairing problem is solved using matching techniques. The authors of~\cite{8632657} consider hybrid OMA and NOMA techniques to solve the problem of maximizing the energy-efficiency while guaranteeing the minimum rate requirements and the maximum transmission power constraints. A swap matching algorithm is proposed to solve the user pairing problem under a fixed power allocation. Once the user pairing is found, the power allocation is solved while maximizing the ratio of the rate to the total power consumption. The authors of~\cite{8365765} study the problem of maximizing the NSD to solve the uplink access problem in NOMA systems as done in our work. Specifically, the objective is to maximize the NSD while allocating the channels to the devices and further guaranteeing the minimum rate requirements and the maximum transmission power constraints. The authors first find the power allocation solution by solving the feasibility problem of the minimum rate requirements. The NOMA channel assignment problem is solved by reducing it to a maximum independent set problem. Nonetheless, the analysis is only given for devices pairing (two devices per NOMA group) and there is no NP-hardness proof, real-time requirements, and power consumption that should be respected to guarantee the long-term operations of the IoT devices. In~\cite{7934461}, resource allocation is considered in the downlink multi-carrier NOMA system to minimize the total transmission power. The problem is formulated as a non-convex programming problem that involves power allocation, rate allocation, user scheduling, and SIC decoding. A brand-and-bound optimal and suboptimal iterative algorithms are proposed to solve the complex problem. In~\cite{7812683}, a multi-carrier NOMA system is considered where full-duplex base stations are employed. The problem that is considered involves the maximization of the weighted sum rate and is formulated as a non-convex programming problem. The monotonic optimization approach is used to develop an optimal algorithm and the successive convex optimization approach is used to design a suboptimal algorithm. In~\cite{8638934}, the authors study the problem of minimizing the maximum access delay under the minimum rate requirements and prove its NP-hardness. They divide the problem into user scheduling and power control subproblems. The user scheduling subproblem is solved using a graph cutting method while the power control subproblem is solved using an iterative algorithm. The authors of~\cite{8764608} study the problem of maximizing the NSD in multi-carrier uplink NOMA systems as done in our work. They proposed a mathematical programming formulation to solve the problem. However, the problem does not include any real-time requirements for the transmission of the IoT devices as done in our problem formulation. Further, their power consumption constraints are defined over the allocated sub-carriers. The power consumption constraints in our work are defined over the frames to guarantee a long-term operation of the IoT devices. Besides,~\cite{8764608} does not study the computational complexity of maximizing the NSD in such an IoT network as done in our work. In a previous work~\cite{8918301}, we studied the problem of user association and scheduling in dense cellular networks and we presented an in-depth characterization of the NP-hardness of the problem. However, the NP-hardness analysis in~\cite{8918301} comes mainly from associating the users to multiple base stations or using multiple channels. Here, we consider the problem of grouping and power allocation for a large number of IoT devices using one channel and one base station. The NP-hardness of the current work, however, comes from (i) the NOMA grouping (ii) the interference structure in NOMA, and (iii) the scheduling of the devices inside the same NOMA group. Further, in this work, we propose an online competitive algorithm in a reinforcement learning framework which was not previously proposed in~\cite{8918301}.

In the previously cited papers, there are some important research gaps that we fill in this work. First, the majority of the works focus on sum-rate related objectives, and thus our work is amongst the fewest to consider maximizing the NSD objective function (e.g.,~\cite{8365765,8764608}). Second, even when the objective is to maximize the NSD, the previously studied problems and our problem are fundamentally different (e.g., no real-time requirements, no long-term power operations, no NP-hardness results). To the best of our knowledge, this work is the first to (1) study the computational complexity of the NSD maximization problem in uplink NOMA IoT networks, (2) propose joint competitive and learning algorithms, and (3) takes into account stringent IoT device requirements including real-time, rate and long-term power requirements.

\subsection{Contributions}\label{sec:contr}
The main contributions of this work are the following.
\begin{itemize}
    \item We model GPA as a mixed-integer non-linear program that can be transformed into an integer linear program for some special cases of interest.
    \item We study the NP-hardness of GPA in different cases. We analyze each case by either presenting a formal proof of NP-hardness or a polynomial time algorithm. 
    \item We start by analyzing GPA in some important special case, where the problem involves only NOMA grouping and scheduling, and we derive online competitive algorithm to solve it. The offline-to-online worst-case ratio of the proposed online competitive algorithm is $2$.
    \item We combine our proposed online competitive algorithm in a machine learning setting in order to solve GPA in the general case. That is, we propose a reinforcement learning algorithm based on the exponential-weighting for exploration and exploitation (\textsc{exp3})~\cite{9013525} method that learns efficiently the power allocation solution based on the online competitive algorithm.
\end{itemize}

The first contribution of this paper is writing GPA as an integer linear programming problem. This helps in solving the problem using off-the-shelf solvers. The second contribution consists of a detailed and in-depth complexity analysis of GPA in IoT networks. The complexity analysis classifies GPA into the polynomial time solvable class and the non-deterministic polynomial time (NP) solvable class. This classification helps in designing efficient algorithms depending on the complexity results, e.g., for NP-hard versions, it is ``useless'' to try to find polynomial time optimal solutions. The next contribution is the design of a new competitive-based reinforcement learning framework to solve the problem. Our framework, first, designs a competitive algorithm to solve a set of simpler subproblems. Then, based on the stochastic shortest path approach and the Markov decision process modelling, the reinforcement learning tools are used to design efficient and close-to-optimal algorithms. The other contributions of our work are related to the problem definition itself: we define a problem that better fits the IoT system model and that aims to maximize the objective function consisting of the NSD. Instead of studying this particular objective function, most related works study the maximization of sum-rate-related or the minimization of sum-power-related objective functions.

In this paper, GPA is solved over a set of frames during which each device must respect its transmission power. To solve this challenging problem, we follow a divide-and-conquer-related framework. Particularly, we divide GPA into several subproblems---one for each frame. Then, for each frame, we propose an optimal online competitive algorithm. To combine the solutions of each subproblem, we propose a learning approach that only learns the power allocation for each IoT device. Note that applying our learning approach to the whole problem from the beginning is also possible but may complicate the learning process because it will involve learning (i) the device grouping (ii) the scheduling decisions over a short period as well as (iii) the power allocation. This exponentially increases the number of states and actions in the learning framework. Further, applying advanced machine learning techniques such as deep reinforcement learning~\cite{9524882,mlika2021network} to solve GPA and deal with the curse of dimensionality might not be scalable as the number of devices becomes massive since deep reinforcement learning require often implementing training-intensive tasks (see remark~\ref{remark:dql}). However, we will investigate the use of deep reinforcement learning to solve GPA for our future work. For this work, to simplify the learning process, we propose a two-fold scheme inspired by the divide and conquer framework. In the first step, we start by solving the problem of device grouping and scheduling in a single frame by proposing a simple and optimal online competitive algorithm. Thereafter, this online algorithm is used as a subroutine (or a black box) in the learning process to find the power allocation for the NOMA groups.

The main idea of the proposed online competitive algorithm comes from the combinatorial online graph matching techniques, e.g.,~\cite{1360462}. Regarding the online learning framework, the main idea of our algorithm is based on \textsc{exp3} and the stochastic shortest path techniques. \textsc{exp3} is a well-known reinforcement learning algorithm that have sub-linear regret~\cite{10.1137/S0097539701398375}.

\subsection{Organization}\label{sec:org}
The rest of the paper is organized as follows. Section~\ref{sec:model} presents the system model. Section~\ref{sec:formulation} formulates GPA as a mathematical program. Section~\ref{sec:NPhard} studies its NP-hardness in different cases and discusses its offline solutions. Section~\ref{sec:onsolutions} presents the proposed online competitive solutions whereas Section~\ref{sec:learn} presents the proposed learning algorithms. Section~\ref{sec:simulations} shows some simulation results and finally, section~\ref{sec:conclusions} draws some conclusions.

\subsection{Notations}
Lowercase and boldface letters denote vectors whereas uppercase and boldface letters denote multi-dimensional arrays. A set of $n$ elements is denoted by $[n]\coloneq\{1,2,\ldots,n\}$ and its cardinality is denoted by $|[n]|\coloneq n$. Interval of integers is denoted by $\{a..b\}$ including $a$ and $b$. The real interval $[0,x]$ is sampled using the integer power level $\tau\geqslant1$ to obtain the set $[x]_{\tau}\coloneq\{\ell x/\tau : \ell = 0,1,\ldots,\tau\}$ with cardinality $|[x]_{\tau}|=\tau+1$. The symbol $\mathscr{O}(\cdot)$ denotes the big-O notation.

\section{System Model}\label{sec:model}
\begin{figure}
\centering
        \includegraphics[scale=0.45]{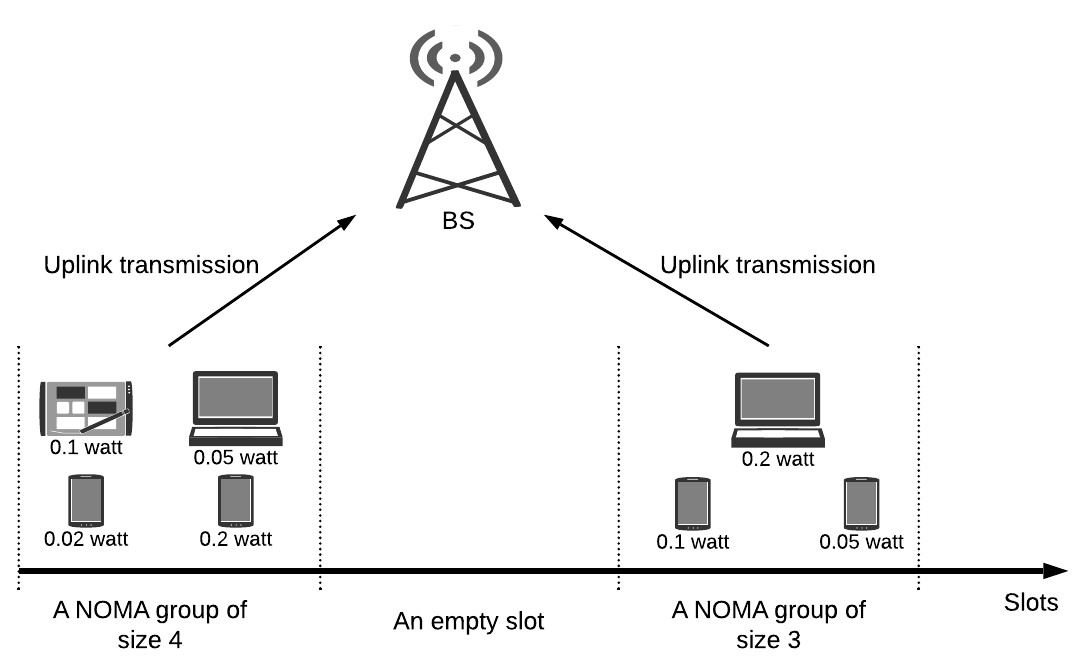}
    \caption{System model with a single frame. Power allocation is shown below each device. An empty slot represents the case of constraint violation (e.g., real-time or power constraints).}
    \label{fig:model}
\end{figure}

Let us consider one base station (BS) and $m$ devices in a cellular-based IoT network. Time is divided into $k$ frames where each frame is composed of $n$ time-slots with unit length each. In each frame, device $i$ may have a packet to send. The length (in bits) of device $i$'s packet in frame $t$ is $L_{i}^t\geqslant0$. (In general, we may have $L_{i}^t=0$ for some $t$; meaning that $i$ has no packet to send in frame $t$.) The arrival time and the deadline of device $i$'s packet in frame $t$ are denoted by $a_{i}^t$ and $d_{i}^t$, respectively. We consider a traffic pattern that is similar to but more general than the well-known frame-synchronized traffic pattern used in~\cite{8482322,8049295}. Device $i$ has $\upper{e}_{i}$ units of energy stored in its battery. For simplicity, a resource block (RB) is represented simply by a time-slot (but time/frequency RBs can also be used). Every frame has $n$ RBs and we denote a RB by the pair $(j, t)$ for time-slot $j$ of frame $t$. A RB has a bandwidth of $W$ Hz. An example of the system model is given in Fig.~\ref{fig:model}. Our system model captures the fast uplink grant in IoT networks and machine-type communications~\cite{8663999} as well as the NOMA uplink massive connectivity scenarios~\cite{8638934}.

The wireless channel between device $i$ and the BS using RB $(j,t)$ is given by $h_{ij}^t$, which may include fast and slow fading. Let $x_{ij}^t$ be a binary variable that is 1 if and only if device $i$ is served using RB $(j,t)$. Also, let $p_{ij}^t$ denotes the transmission power of device $i$ using RB $(j,t)$. Finally, let $z_{i}^t$ be a binary variable that is 1 if and only if device $i$ is served in frame $t$. We use $\mathbf{X}$, $\mathbf{P}$, and $\mathbf{Z}$ to denote the multidimensional variables corresponding to $x_{ij}^t$, $p_{ij}^t$ and $z_{i}^t$, respectively.

The signal to interference-plus-noise ratio (SINR) between device $i$ and the BS using RB $(j,t)$ is given by:
\begin{align}
  \label{sinr}
  \mathit{SINR}_{ij}^t(\mathbf{X},\mathbf{P})=\dfrac{x_{ij}^tp_{ij}^tg_{ij}^t}{1+I_{ij}^t(\mathbf{X},\mathbf{P})},
\end{align}
where $g_{ij}^t=|h_{ij}^t|^2$ is the channel power gain\footnote{We normalize the channel power gain to get a noise power of $1$.} and $I_{ij}^t(\mathbf{X},\mathbf{P})$ is the power of the interference coming from other devices transmitting using RB $(j,t)$. 

To serve a large number of devices, power-domain NOMA technique is used~\cite{7557079}, where a group of devices are transmitting to the BS using the same RB. Successive interference cancellation (SIC) is used at the BS for the decoding. Let $\mathbb{A}_{j}^t$ be the set of devices that are transmitting on RB $(j,t)$. It is well-known to use the highest channel decoding order in uplink NOMA~\cite{7557079,8632657}. In other words, the interference received by the BS, which is generated by the transmission of device $i$'s packet, comes from all devices that have lower channel gains. We order the devices in $\mathbb{A}_{j}^t$ with respect to $g_{ij}^t$ to obtain a new set $\mathbb{B}_{ij}^t\coloneq\{i'\in\mathbb{A}_{j}^t:g_{i'j}^t<g_{ij}^t\}.$ With that said, the interference received by the BS, which is generated by device $i$'s packet using RB $(j,t)$, can be calculated as follows:
\begin{align}\label{interference}
  I_{ij}^t(\mathbf{X},\mathbf{P})=\sum_{i'\in\mathbb{B}_{ij}^t}x_{i'j}^tp_{i'j}^tg_{i'j}^t.
\end{align}

The achievable rate between device $i$ and the BS using RB $(j,t)$ is given by:
\begin{align}\label{rate}
  R_{ij}^t(\mathbf{X},\mathbf{P})=W\lg(1+\text{SINR}_{ij}^t(\mathbf{X},\mathbf{P})).\quad\textit{[in bits/s]}
\end{align}

The objective of GPA is to maximize the number of times the devices are served during the time horizon of $k$ frames. This has to be done while grouping the devices in each RB and allocating their transmission powers. GPA guarantee that the served devices respect (i) their data requirements by sending all of their bits in each frame (ii) their maximum transmission powers, and (iii) their arrival times and deadlines. 

GPA is studied in the \textit{online} scenario with a BS-based centralized implementation as well as a device-based distributed implementation. On the one hand, the BS-based centralized implementation uses competitive analysis and is a slot-based process. During each slot $j$, the BS executes the proposed competitive algorithm and knows only the current and previous (with respect to $j$) information of all devices. This information is acquired by the BS using standard pilot measurements. On the other hand, the device-based distributed implementation uses reinforcement learning and is a frame-based process. At the beginning of each frame $t$, each device uses its current and previous information to select a transmission power according to the proposed learning framework. Next, each device communicates, at the beginning of each frame, its decision to the BS that will execute the competitive algorithm as a black box (see Fig.~\ref{fig:block}). Note that the BS broadcasts its collected information to all devices and thus the channel state information is assumed to be collected for resource allocation using pilot measurements with small signalling overhead.

To solve GPA optimally in the offline scenario using off-the-shelf solvers, we formulate it as a mathematical program in the next section. This optimal offline solution is used to benchmark other proposed algorithms.

\section{Problem Formulation}\label{sec:formulation}

GPA is formulated as follows:
\begin{problem}\label{pb:1}
  \begin{alignat}{2}
  & \maximize_{\mathbf{X},\mathbf{P},\mathbf{Z}} &\qquad &\sum_{i=1}^m\sum_{t=1}^kz_{i}^t\label{obj:1}\\
  & \subjto
  & & x_{ij}^t,z_{i}^t\in\{0,1\},p_{ij}^t\geqslant0,\forall i,j,t,\label{cns1:1}\\
  & & & \sum_{j=1}^nR_{ij}^t(\mathbf{X},\mathbf{P})\geqslant L_{i}^tz_{i}^t,\forall i,t,\label{cns2:1}\\
  & & & p_{ij}^t\leqslant\upper{e}_ix_{ij}^t,\forall i,j,t,\label{cns3:1}\\
  & & & \sum_{j=1}^n\sum_{t=1}^k p_{ij}^t \leqslant\upper{e}_i,\forall i,\label{cns4:1}\\
  & & & x_{ij}^t = 0, \forall i,j\notin\{a_{i}^t..d_{i}^t-1\},t,\label{cns5:1}\\
  & & & x_{ij}^t \leqslant z_{i}^t, \forall i,j,t,\label{cns6:1}\\
  & & & \sum_{i=1}^mx_{ij}^t \leqslant M, \forall j,t,\label{cns7:1}\\
  & & & z_{i}^t\leqslant L_{i}^t, \forall i,t.\label{cns8:1}
  \end{alignat}
\end{problem}
The objective function in~\eqref{obj:1} maximizes the number of times each device is served during the time horizon of $k$ frames. Constraints~\eqref{cns1:1} list the optimization variables. Constraints~\eqref{cns2:1} guarantee a minimum of $L_{i}^t$ bits for device $i$ when served in frame $t$. Constraints~\eqref{cns3:1} guarantee that if device $i$ is not served in slot $j$ of frame $t$, then it is not using any power and it is using at most the maximum otherwise. Constraints~\eqref{cns4:1} restrict the limit of transmission power used by device $i$ in all RBs $(j,t)$. Constraints~\eqref{cns5:1} force $x_{ij}^t=p_{ij}^t=z_{i}^t=0$ for all $i,j,t$, whenever device $i$'s packet has not yet arrived or its deadline is already due. Constraints~\eqref{cns6:1} guarantee that if device $i$ is served in frame $t$, then there must exist a slot $j$ when it is served. Constraints~\eqref{cns7:1} limit the number of devices served in RB $(j,t)$ to a positive number $M\leqslant m$. Finally, constraints~\eqref{cns8:1} mark device $i$ as not-yet-served in frame $t$ if it has no packet to send.

We can see that~\eqref{pb:1} is non-linear and non-convex due to the multiplication of $\mathbf{X}$ and $\mathbf{P}$ in constraints~\eqref{cns2:1}. Note that the variables $\mathbf{X}$ and $\mathbf{P}$ can be related, i.e., $p_{ij}^t>0$ if and only if $x_{ij}^t=1$. Thus, we can get rid of $\mathbf{X}$ from constraints~\eqref{cns2:1} by writing $R_{ij}^t(\mathbf{X},\mathbf{P})=R_{ij}^t(\mathbf{1},\mathbf{P})$. Despite this fact,~\eqref{pb:1} is still MINLP, which is hard to solve in general. 

In the sequel, the transmission power of each device are assumed to belong to some discrete set. This is a realistic assumption in many real systems~\cite{8052127,7010886,4215595}. Although the continuous power assumption can ease mathematical derivations through mathematical programming, GPA is still NP-hard even under the discrete power assumption (as it will be shown shortly). Under the discrete power assumption, every device $i$ can choose its transmission power $p_{ij}^t$ from the set $[\upper{e}_i]_{\tau_i}\coloneq\{0,\upper{e}_i/\tau_i,2\upper{e}_i/\tau_i,\ldots,\upper{e}_i\}$, where $\tau_i\geqslant1$ is the power level of device $i$. An important case called the binary power (BP) case is when $\tau_i=1$ for all $i$, and thus $[\upper{e}_i]_{1}=\{0,\upper{e}_i\}$. We call the general case of $[\upper{e}_i]_{\tau_i}$ the general power (GP) case. 
The BP case is worth studying because it helps understand the intrinsic difficulty of the problem and helps in characterizing the structure of the solution in more general cases.
\begin{remark}[Reduction from multiple frames to a single frame]\label{rmf}
	Note that the devices must be served during the set of $k$ frames while respecting their limited operating energy levels. Thus, in an optimal solution of GPA, every frame will be allocated a feasible amount of power for each device. If one could find how much power to allocate to each device in each frame, then one could reduce the problem to a set of $k$ single frame subproblems and solve each one separately. Therefore, to solve the whole problem over the set of $k$ frames, one must know how to solve the subproblem of a single frame first. To do so, a fixed amount of power allocation is assumed in each frame for each device. Two questions arise: (1) how to solve the single frame subproblem? And, (2) how to allocate the power in each frame for each device? We start by answering the first question (the superscript $t$ is dropped from all the notations). Once each device knows how to solve each subproblem given a fixed power allocation, it has constructed its black box. Next, to answer the second question and thus to find a joint close-to-optimal solution to the whole problem including the power allocation to use in each frame for each device, a reinforcement learning approach is proposed that uses the constructed black box as a subroutine.
\end{remark}

Before going into the details of our proposed solution, in the next section, we study GPA in the offline scenario with a single frame. We give some insights into the offline solutions. Further, we study its computational complexity by characterizing its NP-hardness in different cases.
\section{NP-hardness and the Offline Scenario}\label{sec:NPhard} 
\subsection{The Offline Problem in the BP Case}\label{sec:offsolutions}
We consider the offline version of GPA for $M=1$. In this case, GPA is equivalent to the following: maximize the NSD during $n$ slots subject to the constraints of arrival times, deadlines, data requirements, and matching capacity (i.e., no more than one device in the same slot and no more than one slot for each device). We can solve this problem by reducing it to a maximum matching problem in a bipartite graph. First, we create a bipartite graph where the slots represent the left vertexes and the devices represent the right vertexes. An edge exists between slot $j$ and device $i$ if and only if $\upper{e}_ig_{ij}\geqslant(2^{L_i/W}-1)$ and $j\in\{a_i..d_i-1\}$. Every edge in the graph has capacity $1$. By introducing a source vertex and sink vertex, we can find the optimal solution to this maximum matching problem in polynomial time by applying some known maximum flow algorithm.

Solving GPA for general $M>1$ seems to be a hard task. Indeed, we show in the following that GPA is NP-hard for any fixed $M\geqslant3$. Nonetheless, it remains open whether or not GPA in the BP case is NP-hard for $M=2$.
\begin{theorem}
    GPA is NP-hard in the BP case for $M\geqslant3$.
\end{theorem}

\begin{IEEEproof}
We reduce 3-bounded 3-dimensional matching (3DM3)~\cite{Garey:1979} to GPA. In 3DM3, we have a set $\mathbb{T}\subseteq\mathbb{W}\times\mathbb{X}\times\mathbb{Y}$, where $\mathbb{W}$, $\mathbb{X}$, and $\mathbb{Y}$ are disjoint sets having the same number $\ell$ of elements and $|\mathbb{T}|=r$. Also, in 3DM3, no element of $\mathbb{W}\cup\mathbb{X}\cup\mathbb{Y}$ occurs in more than three triples of $\mathbb{T}$. We may assume without loss of generality that $\ell<r<2\ell$. The goal is to find in $\mathbb{T}$ a matching of a maximum size, i.e., a subset $\mathbb{M}\subseteq\mathbb{T}$ where no two elements of $\mathbb{M}$ agree in any coordinate.

Given an instance of 3DM3, an instance of GPA is obtained as follows. Let $M=3$. We create $n=r$ slots; slot $j$ corresponds to the 3-element set $t_j$ from $\mathbb{T}$. We also create $m=2\ell+r$ devices. Specifically, for each $i\in\mathbb{W}$, we have a device $w_i$, for each $i\in\mathbb{X}$, we have a device $x_i$, and for each $i\in\mathbb{Y}$, we have a device $y_i$. Also, there are $r-\ell$ additional devices $\{z_1,z_2,\ldots,z_{r-\ell}\}$. Let $a_{i}=1$ and $d_{i}=n$, that is device $i$'s packet arrived at the beginning of the frame and is due at its end. For all devices $i$, set $\upper{e}_i=1$ and let $\Delta$ be a large number. For each device $w_i$ and slot $j$ corresponding to the 3-element set $t_j$, let $b_{w_i}=1$ and
\begin{align*}
    g_{w_ij}\coloneq
    \begin{cases}
    3, & \text{if $i\in t_j$}, \\
    2+w_i/\Delta, & \text{otherwise}.
    \end{cases}
\end{align*}
For each device $x_i$ and slot $j$ corresponding to the 3-element set $t_j$, let $b_{x_i}=1/2$ and
\begin{align*}
    g_{x_ij}\coloneq
    \begin{cases}
    1, & \text{if $i\in t_j$}, \\
    2+x_i/\Delta, & \text{otherwise}.
    \end{cases}
\end{align*}
For each device $y_i$ and slot $j$ corresponding to the 3-element set $t_j$, let $b_{y_i}=1/2$ and
\begin{align*}
    g_{y_ij}\coloneq
    \begin{cases}
    1/2, & \text{if $i\in t_j$}, \\
    2+y_i/\Delta, & \text{otherwise}.
    \end{cases}
\end{align*}
And for each additional device $z_i$, and slot $j$, let $b_{z_i}=1$ and $g_{z_ij}=2+z_i/\Delta$. This instance is clearly created in polynomial time. We prove that 3DM3 is solved with a matching of size $\ell$ if and only if GPA is solved with $2\ell+r$ served devices and each slot serves at most $3$ devices.

On the one hand, if 3DM3 is solved, then for each 3-element set $m_j$ of the matching $\mathbb{M}$, we can serve three devices in slot $j$---one for each element of $m_j$. This is valid because each element of $m_j$ comes, respectively, from $\mathbb{W}$, $\mathbb{X}$, and $\mathbb{Y}$ and thus the corresponding three devices have channel gains equal, respectively, to $3$, $1$, or $1/2$. This means that, for the device $w_i$ we have $3/(1+1+1/2)=6/5\geqslant1$, for the device $x_i$ we have $1/(1+1/2)=2/3\geqslant1/2$ and for the device $y_i$ we have $1/2\geqslant1/2$. We conclude that the three devices meet their data requirements. Since $\mathbb{M}$ is a matching in $\mathbb{T}$ of size $\ell$, the slots corresponding to $\mathbb{M}$ serves a total of $3\ell$ devices---three devices in each slot. The remaining slots can serve at most $(r-\ell)$---at most one device per slot. To maximize the number of devices served, each remaining slot can serve exactly one device. Thus, there are a total of $3\ell+r-\ell=2\ell+r$ devices served where each slot serves at most three devices.

On the other hand, assume that GPA is solved where each slot serves at most three devices with a total of $2\ell+r$ served devices. We argue that if a slot serves three devices, then these devices correspond to some triple in $\mathbb{T}$ (they have channel gains $3$, $1$ or $1/2$). Thus, all slots that serve exactly three devices correspond to a matching in $\mathbb{T}$. Note that, in the solution to GPA, we cannot serve two devices in each slot (for a total of $2r$ devices) because $2\ell+r>2r$. Thus, to maximize the number of devices served, $\ell$ slots need to serve three devices (and $r-\ell$ slots serves one device each), which corresponds to a matching in $\mathbb{T}$. 

The reduction is clearly done in polynomial time. Finally, since 3DM3 is NP-hard~\cite{Garey:1979}, the theorem follows.
\end{IEEEproof}

\begin{table*}
  \centering
  \setlength\tabcolsep{0pt}
  \caption{Complexity classification}
  \label{tab:np}
  \begin{tabular}{|p{3.3cm}|M{3.5cm}|M{3.7cm}|M{3cm}|M{3cm}|}\hline
    & GPA with GP & \multicolumn{3}{c|}{GPA with BP}\\\hline
    Group Size & $M=1$ (no NOMA) & $M=1$ (no NOMA) & $M=2$ (NOMA) & $M\geqslant3$ (NOMA)\\\hline
    Complexity Class & NP-hard & Poly & Open? & NP-hard\\ \hline
    Online Algorithm & Open? & $1.58$-competitive~\cite{1360462} & $2$-competitive & $2$-competitive\\  \hline
  \end{tabular}
\end{table*}

\subsection{The Offline Problem in the GP Case}
In the following, we consider the GP case and we prove that GPA is NP-hard even when $M=1$, i.e., in the OMA case.
\begin{theorem}\label{theorem1}
  GPA is NP-hard in the GP case even for $M=1$.
\end{theorem}

\begin{IEEEproof}
  The proof is to show that a special case of GPA is NP-hard. Let $M=1$. Also, let $a_{i}=1$ and $d_{i}=n$, that is device's $i$ packet arrived at the beginning of the frame and is due at its end. Let us assume that $\upper{e}_i$ is large enough so that device $i$ cannot deplete its energy. Device $i$ is transmitting with fixed power $p_{ij}$ such that $\sum_{j=1}^np_{ij}\leqslant\upper{e}_i$. Denote by $G_{ij}\coloneq\lg(1+p_{ij}g_{ij})$. 

Under this restriction, we prove that GPA is NP-hard using a reduction from the maximum independent set (MIS) problem in graph theory. MIS is defined~\cite{Garey:1979} as follows: given a graph and a positive integer $\zeta$. Is there an independent set in the graph of size $\zeta$ or more? \textit{An independent set is a set of vertexes that do not share any edge.} On the other hand, the restricted version of GPA can be recast as: given the coefficients $G_{ij}$ and the size of the packets $L_i$, is there a scheduling of more than $\sigma$ devices, denoted by the set $\mathbb{O}$, such that a slot is used by at most one device and $\sum_{j\in\mathbb{O}}G_{ij}\geqslant L_i?$ 

Given an instance of MIS, we create an instance of GPA in polynomial time as follows: the vertexes are the devices and the edges are the slots. The edges are numbered as $1,2,\ldots,n$. There is an edge between two devices if and only if one of them can be served at that slot. Let $\mathbb{O}_i$ be the set of slots that device $i$ can be served at. Let $L_i\coloneq|\mathbb{O}_i|$ for each device $i$ and set $\zeta=\sigma$. The coefficients $G_{ie}$ for device $i$ and slot $e$ are given by: 
\begin{align}\label{Gij}
  G_{ie}\coloneq
  \begin{cases}
    0, & \text{if $e\notin\mathbb{O}_i$}, \\
    1, & \text{if $e\in\mathbb{O}_i$}.
  \end{cases}
\end{align}
This reduction is clearly done in polynomial time. Now it remains to prove that: there are more than $\zeta$ served devices, denoted by the set $\mathbb{O}$, such that a slot is used by at most one device and $\sum_{j\in\mathbb{O}}G_{ij}\geqslant L_i$ if and only if there is an independent set in the graph of size $\zeta$ or more.

On the one hand, assume that we have an independent set in the graph of size $\zeta$ or more. By setting $x_{ie}=1$ for all $i$ in the independent set and $e\in\mathbb{O}_i$, we have more than $\zeta$ devices served. Further, since we have an independent set, it is true that the served devices are not overlapping with one another.

On the other hand, assume that we have a solution to the restricted version of GPA, then we can see, by construction, that for device $i$ to be satisfied, it must be scheduled in all slots in $\mathbb{O}_i$ (since $L_i=|\mathbb{O}_i|$ and $G_{ie}$ is binary). Since we have more than $\zeta$ non-overlapping served devices, they form, in the corresponding graph, an independent set of size more than $\zeta$.

Summarizing, we have reduced MIS to the restricted version of GPA in polynomial time such that MIS is solved if and only if GPA is solved. Since MIS is a well-known NP-hard~\cite{Garey:1979}, so is GPA. This proves the theorem.
\end{IEEEproof}

The NP-hardness results and the proposed algorithmic solutions are summarized in table~\ref{tab:np}. We use ``Poly'' to denote the polynomial time complexity class and ``Open?'' to denote that, to the best of our knowledge, the problem is still open.

\section{Competitive Algorithms}\label{sec:onsolutions} 
As previously discussed in remark~\ref{rmf}, we start by solving GPA in the single frame case. Then, we solve it in the more general case of multiple frames. Before going into the details, we give the following definition.
\begin{definition}[$c$-competitive algorithm~\cite{Borodin1998}]\ \\
    An online algorithm \textsc{alg} is $c$\textbf{-competitive} if there is a constant $\alpha$ such that for all finite input sequences, $O\leq cA+\alpha$,
    where $A$ (or $O$) is the value returned by \textsc{alg} (or an optimal algorithm) for a given input. \textsc{alg} is \textbf{strictly} $c$\textbf{-competitive} if it is $c$-competitive and $\alpha\leq0$.
\end{definition}

We assume that device $i$ can choose its transmission power from $[\upper{e}_i]_1=\{0,\upper{e}_i\}$. That is, device $i$ can either transmit in a slot or stay silent once and forever during the frame. This assumption implies that a device can use at most one RB. This is realistic in massive IoT networks where devices normally have short data packets to send~\cite{7165674,6786066} and thus require no more than a single RB for transmission.

\begin{remark}[The selfish algorithm]\label{selfish}
Assume that $g_{1j}<g_{2j}<\cdots<g_{mj}$. According to the decoding order of largest channel gains in uplink NOMA, device $1$ knows that if $\upper{e}_1g_{1j}<2^{L_1/W}-1$, then $p_{1j}=0$. Now, if $\upper{e}_1g_{1j}\geqslant2^{L_1/W}-1$, then setting $p_{1j}=\upper{e}_1$ would satisfy device 1 but may interfere with other devices. The following is called the \textit{selfish} algorithm for device $i$: whenever $\upper{e}_ig_{ij}\geqslant2^{L_i/W}-1$, set $p_{ij}=\upper{e}_i$. We can prove that the selfish algorithm can perform very badly. Say device 1 acts selfishly. Then, we can find an instance in which only device 1 will be served in slot $j$ (due to the severe interference that it generates)---removing device 1 from slot $j$ would satisfy all other devices in that slot. Assume we have $L_i=\lg(1+i)$ for all $i$ and $g_{1j}=1$ but for all $i\ne1$, $g_{ij}=\sqrt{2^{(i-2)(i+1)}}\bigl(2^i-1\bigr).$ In this case, if device 1 transmits in slot $j$, then no other device can transmit in that slot. However, if it does not, then all device $i\ne1$ can transmit. This shows that the selfish algorithm has a competitive ratio of at least $m-1$ which is very large.
\end{remark}

The previous remark proves that acting selfishly is not very good in terms of maximizing the NSD. To provide better results, we first study the case of $M=1$ and then generalize the analysis to larger $M$.  

For $M=1$, we can reduce GPA to an online matching problem in a bipartite graph as follows. The devices represent the right vertexes of the bipartite graph. The slots represent the left vertexes that appear online one-by-one. An edge exists between slot $j$ and device $i$ if and only if $\upper{e}_ig_{ij}\geqslant(2^{L_i/W}-1)$ and $j\in\{a_i..d_i-1\}$. When slot $j$ appears, the channel gain $g_{ij}$ is revealed for all devices $i$ and thus the edges incident to it are also revealed. Once revealed, an online algorithm must make an irrevocable decision of which device to serve at slot $j$ (i.e., match the corresponding edge). This online matching problem can be solved using the well-known \textit{ranking} algorithm that has a competitive ratio of $\frac{e}{e-1}\approx1.58$~\cite{1360462}. The ranking algorithm chooses a random permutation $\rho$ of the devices. For each slot $j$, it finds the set of not-yet-served devices $\mathbb{Y}_j$ that can transmit in this slot, i.e., those devices $i$ that have $\upper{e}_ig_{ij}\geqslant(2^{L_i/W}-1)$ and $j\in\{a_i..d_i-1\}$. If $\mathbb{Y}_j$ is not empty, then the ranking algorithm chooses a device $i$ from $\mathbb{Y}_j$ that minimizes $\rho(i)$. The ranking algorithm is equivalent to assigning priorities to the devices and choosing the not-yet-served device that has the highest priority.

To solve the problem for general $M\geqslant1$, we transform it into a many-to-one matching problem and we adopt a greedy approach to solve it. We create the previous same bipartite graph. Now, contrary to the case of $M=1$, each slot can be matched to at most $M$ devices from those connected to it by an edge. For each slot $j\in[n]$, let $\mathbb{N}_j$ denotes the set of neighbours of $j$ (i.e., $\mathbb{N}_j\coloneq\{i\in[m]:\{i,j\} \text{ is an edge}\}$). Once slot $j$ is revealed, the problem is reduced to  finding a set of (at most $M$) devices $\mathbb{D}_j\subseteq\mathbb{N}_j$ of maximum cardinality such that:
\begin{align}\label{binslot}
\upper{e}_ig_{ij}\geqslant\left(2^{L_i/W}-1\right)\Bigl(1+\sum_{i'\in\mathbb{D}_j'}\upper{e}_{i'}g_{i'j}\Bigr),
\end{align}
is valid for each $i\in\mathbb{D}_j$, where $\mathbb{D}_j'\coloneq\{i'\in\mathbb{D}_j:g_{ij}>g_{i'j}\}$.

It is known that the complexity of SIC decoding increases as the number of users transmitting on the same RB increases~\cite{Tse:2005:FWC:1111206}. Thus, in general, $M$ is chosen small in order to keep the complexity of SIC decoding low. In fact, multiple research papers consider the case of user pairing when $M=2$~\cite{8365765,8320533}. Hence, for small and fixed $M$, one could generate, in slot $j$, all combinations of at most $M$ devices and matches the maximum-cardinality set $\mathbb{D}_j$ that respects~\eqref{binslot}. This leads to a polynomial time (only for fixed $M$) worst-case complexity of $\mathscr{O}(m^M)$. However, by analyzing the problem's structure based on~\eqref{binslot}, we provide an optimal way of finding a maximum cardinality set that satisfies~\eqref{binslot} in $\mathscr{O}(m)$ worst-case time complexity.
\begin{lemma}\label{lm:binslot}
  Once slot $j$ is revealed, finding a maximum cardinality set $\mathbb{D}_j$ that satisfies~\eqref{binslot} can be done in $\mathscr{O}(m)$ worst-case time complexity.
\end{lemma}

\begin{IEEEproof}
  We prove that the greedy algorithm, given in Algorithm~\ref{alg:bmj} below and called binary-matching-j ($\textsc{bm}_j$), which is applied at slot $j$, gives a maximum cardinality set that satisfies~\eqref{binslot} in $\mathscr{O}(m)$ time in the worst-case; assuming that the channel gain $\mathbf{g}_j$ is sorted, otherwise the complexity would be $\mathscr{O}(m\lg m)$. 
  \begin{algorithm}[ht!]
  \caption{The $\textsc{bm}_j$ algorithm}
  \label{alg:bmj}
  \begin{algorithmic}[1]
    \Require{$M,m,\mathbb{N}_j,\mathbf{g}_j,\mathbf{L},\upper{\mathbf{e}}$}
    \Ensure{$\mathbb{D}_j$}
    \State $\mathbb{X}\gets\emptyset$
    \State $X\gets0$
    \For{$i\gets1$ \textbf{to} $m$}
        \If{$i$ \textbf{in} $\mathbb{N}_j$}
            \If{$\upper{e}_ig_{ij}\geqslant (2^{L_i/W}-1)(1+X)$}
                \State $\mathbb{X}\gets\mathbb{X}\cup\{i\}$
                \State $X\gets X+\upper{e}_ig_{ij}$
            \EndIf
        \EndIf
    \EndFor
    \If{$|\mathbb{X}|\leqslant M$} $\mathbb{D}_j\gets\mathbb{X}$
    \Else \text{ } Let $\mathbb{D}_j\subset\mathbb{X}$ with $|\mathbb{D}_j|=M$
    \EndIf
    \State\Return $\mathbb{D}_j$
  \end{algorithmic}
\end{algorithm}

The worst-case time complexity of $\textsc{bm}_j$ is clearly $\mathscr{O}(m)$. It remains to show that the algorithm returns a feasible solution of maximum cardinality that satisfies~\eqref{binslot}.

First, it is clear that $|\mathbb{D}_j|\leqslant M$. Using mathematical induction on each iteration of the algorithm, we prove that the set $\mathbb{D}_j$ represents a feasible solution that satisfies~\eqref{binslot}. Let $\mathbb{D}_j^p$ be the set of devices returned by $\textsc{bm}_j$ after iteration $p$. For $p=1$, device $1$ is added to $\mathbb{D}_j^1$ only if $\upper{e}_1g_{1j}\geqslant b_1$ and thus, $\mathbb{D}_j^1$ is feasible. Assume that $\mathbb{D}_j^p$ is feasible. Is $\mathbb{D}_j^{p+1}$ feasible? At iteration $p+1$, the algorithm adds device $p+1$ to $\mathbb{D}_j^{p+1}$ only if $\upper{e}_{p+1}g_{p+1j}\geqslant b_{p+1}(1+X)$ where $X=\sum_{i\in\mathbb{D}_j^p}g_{ij}$. If this condition is not met, then $\mathbb{D}_j^{p+1}=\mathbb{D}_j^{p}$ and we are done. Otherwise, $\mathbb{D}_j^{p+1}=\mathbb{D}_j^{p}\cup\{p+1\}$. Since the channel gains are sorted in increasing order, thus $g_{p+1j}\geqslant g_{ij}$ for all $i\in\mathbb{D}_j^p$. According to the largest channel gain decoding order of SIC, the devices already in $\mathbb{D}_j^{p}$ will not be affected by the transmission of device $p+1$. Because device $p+1$ is added to $\mathbb{D}_j^{p}$ only if~\eqref{binslot} is respected, we conclude that $\mathbb{D}_j^{p+1}$ is feasible. Combining the base case and the inductive hypothesis, we finally have that the returned set $\mathbb{D}_j=\mathbb{D}_j^{m}$ is feasible.

To prove the optimality, let $\{i_1,i_2,\ldots,i_{\ell_1}\}$ be the set of served devices in the order they were added to $\mathbb{D}_j$ and let $\{i_1^*,i_2^*,\ldots,i_{\ell_2}^*\}$ be the set of devices in the order they were added to $\mathbb{O}_j$ returned by some optimal algorithm. We assume without loss of generality that $g_{i_1j}<g_{i_2j}<\cdots<g_{i_{\ell_1}j}$ and $g_{i_1^*j}<g_{i_2^*j}<\cdots<g_{i_{\ell_2}^*j}$. The optimality is to prove that $\ell_2=\ell_1$. 

First, we prove by induction on $\ell\leqslant\ell_1$ that $g_{i_\ell^*j}\geqslant g_{i_\ell j}$. The base case, for $\ell=1$, is clearly true: the first device served by $\textsc{bm}_j$ has the smallest channel gain. Assume now that for $\ell>1$ the statement is true for $1,2,\ldots,\ell-1$, i.e., $g_{i_{\ell-1}^*j}\geqslant g_{i_{\ell-1}j}$, is it true for $\ell$? If $g_{i_{\ell}^*j}<g_{i_{\ell}j}$, then $\textsc{bm}_j$ would have chosen $i_{\ell}^*$ instead of $i_{\ell}$ because, using the inductive hypothesis, $g_{i_{\ell}j}>g_{i_{\ell}^*j}\geqslant b_{i_{\ell}^*}(1+\sum_{i=i_{1}^*}^{i_{\ell-1}^*}g_{ij})\geqslant b_{i_{\ell}^*}(1+\sum_{i=i_{1}}^{i_{\ell-1}}g_{ij})$. Thus, for all $\ell\leqslant\ell_1$, it is true that $g_{i_\ell^*j}\geqslant g_{i_\ell j}$.

Now, we use the previous fact to prove, by contradiction, that $\ell_1=\ell_2$. Assume that $\ell_2>\ell_1$. In other words, there exists a device $i_{\ell_1+1}^*\in\mathbb{O}_j$, or equivalently, the optimal algorithm chooses $i_{\ell_1+1}^*$ in iteration $\ell_1+1$. Thus, $g_{i_{\ell_1+1}^*j}\geqslant b_{i_{\ell_1+1}^*}(1+\sum_{i=i_{1}^*}^{i_{\ell_1}^*}g_{ij})\geqslant b_{i_{\ell_1+1}^*}(1+\sum_{i=i_{1}}^{i_{\ell_1}}g_{ij})$, where the last inequality follows form the previous fact. Since the optimal algorithm chooses $i_{\ell_1+1}^*$ in iteration $\ell_1+1$, then $g_{i_{\ell_1+1}^*j}>g_{i_{\ell_1}^*j}\geqslant g_{i_{\ell_1}j}$. We can see that, in iteration $\ell_1+1$, device $i_{\ell_1+1}^*$ has larger channel gain than device $i_{\ell_1}$ and can be added to $\mathbb{D}_j$. Since $\textsc{bm}_j$ stopped adding devices at iteration $\ell_1$, we reach a contradiction and we conclude that $\ell_1=\ell_2$. 

Finally, the set $\mathbb{D}_j$ returned by $\textsc{bm}_j$ is of maximum cardinality and is obtained in $\mathscr{O}(m)$ worst-case time complexity. This proves the lemma.
\end{IEEEproof}
 
The proposed algorithm to solve GPA is called the binary-matching-slots (\textsc{bms}) algorithm and its pseudo-code is given in Algorithm~\ref{alg:onmatchingM}. For each arriving slot, \textsc{bms} calls $\textsc{mb}_j$ and serves the maximum possible number of devices in that slot. Then, it updates the set of not-yet-served devices and continues in this way. We prove that this algorithm is $2$-competitive. 
\begin{algorithm}[ht!]
  \caption{The \textsc{bms} algorithm}
  \label{alg:onmatchingM}
  \begin{algorithmic}[1]
    \Require{Bipartite graph, $M,m,n,\mathbf{g},\mathbf{L},\upper{\mathbf{e}}$}
    \Ensure{$\{\mathbb{D}_1,\mathbb{D}_2,\ldots,\mathbb{D}_n\}$}
    \State $\mathbb{X}\gets[m]$
    \For{each slot $j$}
        \State $\mathbb{D}_j\gets\textsc{bm}_j(M,m,\mathbb{N}_j,\mathbf{g}_j,\mathbf{L},\upper{\mathbf{e}})$%
        \State $\mathbb{X}\gets\mathbb{X}\backslash\mathbb{D}_j$
    \EndFor
    \State\Return $\{\mathbb{D}_1,\mathbb{D}_2,\ldots,\mathbb{D}_n\}$
  \end{algorithmic}
\end{algorithm}

\begin{theorem}\label{the2com}
    \textsc{bms} is $2$-competitive.
\end{theorem}

\begin{IEEEproof}
    Let $\mathbb{D}_M\coloneq\{\mathbb{D}_1,\mathbb{D}_2,\ldots,\mathbb{D}_n\}$ be the set of devices served by \textsc{bms} and let $\mathbb{O}_M=\{\mathbb{O}_1,\mathbb{O}_2,\ldots,\mathbb{O}_n\}$ be the set of devices served by some optimal algorithm \textsc{opt}, where $\mathbb{D}_j$ (resp. $\mathbb{O}_j$) is the set of the devices served by \textsc{bms} (resp. by \textsc{opt}) at slot $j$.

    Based on lemma~\ref{lm:binslot}, it is clear that the number of devices in $\mathbb{O}_j\backslash\mathbb{D}_M$ is at most the number of devices in $\mathbb{D}_j$; since otherwise \textsc{bms} would have chosen $\mathbb{O}_j\backslash\mathbb{D}_M$ instead of $\mathbb{D}_j$. Thus, by summing over $j$, the number of devices in $\mathbb{O}_M\backslash\mathbb{D}_M$ is at most the number of devices in $\mathbb{D}_M$. Since $\mathbb{O}_M\subseteq(\mathbb{D}_M\cup\mathbb{O}_M\backslash\mathbb{D}_M)$, thus we obtain:
    \begin{align*}
        |\mathbb{O}_M|&\leqslant|\mathbb{D}_M|+|\mathbb{O}_M\backslash\mathbb{D}_M|,\\
                    &\leqslant|\mathbb{D}_M|+|\mathbb{D}_M|\leqslant2|\mathbb{D}_M|.
    \end{align*}
\end{IEEEproof}

\begin{theorem}\label{theo:mub}
    There is no deterministic online algorithm with better competitive ratio than \textsc{bms}.
\end{theorem}
\begin{IEEEproof}
    Assume $m=n=2$. Let $\upper{e}_1=\upper{e}_2=1$ and $L_1=L_2=1$. The channel gains in slot 1 is $\mathbf{g}_1=[1,1]$. Now, if an online algorithm decides to serve device 1 (resp. 2) in slot 1, then we can choose the channel gains in slot 2 as $\mathbf{g}_2=[1,0]$ (resp. $\mathbf{g}_2=[0,1]$). An optimal offline algorithm can serve device 2 in slot 1 and device 1 in slot 2 if $\mathbf{g}_2=[1,0]$ or it can serve device 1 in slot 1 and device 2 in slot 2 if $\mathbf{g}_2=[0,1]$. In any case, the offline-to-online ratio is $2$.
\end{IEEEproof}

Theorems~\ref{the2com} and~\ref{theo:mub} show that, in the worst-case, the gap between the value of the solution given by \textsc{bms} and the value of the solution given by an optimal omniscient offline algorithm is $2$. This means that, in the worst-case, \textsc{bms} gives at least $50\%$ of the NSD compared to what an optimal omniscient offline algorithm gives.

\subsection{Benchmark Algorithms}\label{SubSec1}
For comparison purposes, in this section, we present an adapted version of a clustering algorithm, called hereinafter \textsc{ath}, by Ali, Tabassum, and Hossain~\cite{7557079}. The original algorithm is offline and works with channel gains that are independent of the RBs. We transform it into an online algorithm as follows. First, for simplicity, here, we assume that $m$ is a multiple of $M$ and denote by $\kappa\coloneq m/M$. For each new slot $j$, \textsc{ath} creates $\kappa$ clusters where each cluster contains exactly $M$ devices by sorting the channel gains in descending order. That is, if $g_{1j}>g_{2j}>\cdots>g_{mj}$, then cluster $l$ will contain the devices $\{l,\kappa+l,2\kappa+l,\ldots,M\}$. Now, for each slot $j$, \textsc{ath} iterates the clusters and checks whether the arrival times, deadlines, and the data rate requirements of the devices in cluster $l$ are respected. If not, the devices are removed iteratively from cluster $l$ until the constraints are not violated. Once all clusters are checked, \textsc{ath} picks the cluster with the maximum NSD. For subsequent slots, \textsc{ath} proceeds similarly with the exception that an already served device is removed from the clusters.

We present another adapted version of a benchmark algorithm, called \textsc{zz}, by Zhai and Zhang~\cite{8365765}. The original algorithm is offline, based on solving independent sets in graphs, and proposed only for $M=2$~\cite{8365765}. The modified version, \textsc{zz}, works as follows. Since $M=2$, it generates all pairs of devices in each slot. By checking the constraints of the arrival time, the deadlines, and the data rate requirements, the pairs are updated in each slot---meaning that a pair can be reduced to a single element or to empty if necessary. \textsc{zz} constructs an undirected graph $G$ where the set of nodes is the possible set of devices (paired or not) in each slot. So, a node $v$ is given by the tuple $(c,j)$ where $c$ is either a single device or a pair of devices served in slot $j$. An edge between node $(c,j)$ and node $(c',j')$ exists if and only if $j=j'$ or $c\cap c'$ is not empty. Once the graph is constructed, \textsc{zz} creates a new graph $H$ by splitting every node $(c,j)$ in $G$ with $|c|=2$ into two nodes that has the same neighbours as in $G$ but are not linked by an edge~\cite{8365765}. Device pairing is now obtained by solving the problem of the maximum independent set in the new graph $H$ using a greedy approach. \textsc{zz} is still offline since it must construct the graph $G$ by knowing all the upcoming slots. 

\subsection{Running Time Complexity}
Here, we analyze the worst-case time complexities of the different algorithms. We summarize the results in table~\ref{tab:time}. The complexity of \textsc{bms} is clearly $\mathscr{O}(nf(m))$  where $\mathscr{O}(f(m))$ is the complexity of $\textsc{bm}_j$ which is equal to $\mathscr{O}(m)$ for sorted channel gains $\mathbf{g}_j$ or $\mathscr{O}(m\lg m)$ otherwise. Similar analysis can be done to obtain $\mathscr{O}(nf(m))$ complexity for \textsc{ath}. As for \textsc{zz}\footnote{As discussed previously, \textsc{zz} is proposed only for user pairing, i.e., $M=2$.}, the generation of all pairs is done in $\mathscr{O}(m^2)$. To construct the graph, one has to iterate the slots and this gives a complexity of $\mathscr{O}(nm^2)$. Once the graph is constructed, finding a maximal independent set using the classical greedy approach requires $\mathscr{O}(m^4)$ complexity since the graph has $\mathscr{O}(m^2)$ nodes. The complexity of the optimal algorithm, denoted \textsc{opt}, obtained by solving~\eqref{pb:1} using off-the-shelf solvers is exponential the worst-case case~\cite{8918301}.
\begin{table}[htb!]
  \centering
  \setlength\tabcolsep{0pt}
  \caption{Worst-case time complexities}
  \label{tab:time}
  \begin{tabular}{|M{3cm}|M{4cm}|}\hline
    Algorithms & Complexity\\\hline
    \textsc{bms} & $\mathscr{O}(nm\lg m)$ \\\hline
    \textsc{ath} & $\mathscr{O}(nm\lg m)$ \\\hline
    \textsc{zz} & $\mathscr{O}(nm^2+m^4)$\\\hline
    \textsc{opt} & Exponential\\\hline
  \end{tabular}
\end{table}

\section{Learning Algorithms}\label{sec:learn}
When there are multiple frames, without further assumption, one cannot hope to obtain good performances in terms of competitiveness. Specifically, say there are two frames and a single device. If an online algorithm decided to allocate some transmission power $p>0$ in frame one. Then, an adversary can always choose the channel gains such that $p\max_j\{g_{j}\}<2^{L/W}-1$ but $p'g_{j}\geqslant2^{L/W}-1$ for some slot $j$ with $p'>p$. Next, the adversary can also choose the channel gains in the second frame such that $\upper{e}\max_j\{g_{j}\}<2^{L/W}-1$. In this manner, the adversary can serve the device once in frame one with $p'$ but an online algorithm never served the device. For this reason, we are motivated to consider a relative performance measure and thus we adopt the learning framework. 

To obtain a global solution in the multi-frame model to GPA, we use machine learning techniques. Specifically, we combine \textsc{bms} and reinforcement learning~\cite{amine} techniques to obtain the NOMA grouping and scheduling as well as the power allocation solutions.

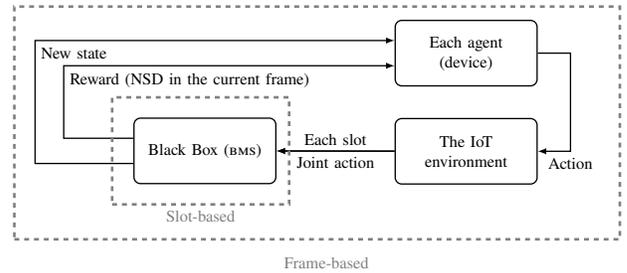
\begin{figure}[ht!]
  \centering
  \resizebox{.45\textwidth}{!}{%
\begin{tikzpicture}[node distance = 6em, auto, thick]
    \node [rectangle,draw,text width=8em,text centered,rounded corners,minimum height=4em] (Agent) {Each agent (device)};
    \node [rectangle,draw,text width=8em,text centered,rounded corners,minimum height=4em, below of=Agent] (Environment) {The IoT environment};
    \node [rectangle,draw,text width=8em,text centered,rounded corners,minimum height=4em, left of=Environment] at (-3.5, -2.1) (BlackBox) {Black Box (\textsc{bms})};
    
     \path [draw, -latex] (Agent.0) --++ (2em, 0em) |- node []{Action} (Environment.0);
     \path [draw, -latex] (Environment) --++ (BlackBox) node [midway, above]{Each slot};
     \path [draw, -latex] (Environment) --++ (BlackBox) node [midway, below]{Joint action};
     \path [draw, -latex] (BlackBox.190) --++ (-6em,0em) |- node [pos=0.45,right] {New state} (Agent.170);
     \path [draw, -latex] (BlackBox.170) --++ (-4.25em,0em) |- node [pos=0.4,right] {Reward (NSD in the current frame)} (Agent.190);
     \node (rect) at (-5.7,-2.1) [draw,dashed,color=gray,ultra thick,minimum width=3.8cm,minimum height=2.3cm] {};
     \node[gray,ultra thick] at (-5.7,-3.5) {Slot-based};
     \node (rect) at (-3.2,-1.5) [draw,dashed,color=gray,ultra thick,minimum width=13cm,minimum height=5cm] {};
     \node[gray,ultra thick] at (-3,-4.5) {Frame-based};
\end{tikzpicture}
  }
  \caption{The system block of the learning framework.}
  \label{fig:block}
\end{figure}
In Fig.~\ref{fig:block}, we draw the system block of our proposed learning framework to solve GPA. Each agent (or each device) interacts independently with the IoT environment and takes actions accordingly. Specifically, the learning is a frame-based process. In each frame $t$, each device $i$ observes the IoT environment and chooses a transmission power $e_i^t$ from its available set of actions. Once all devices choose their transmission powers, a joint action if formed and a slot-based process is invoked---the proposed \textsc{bms} algorithm---as a black box. Just before the beginning of the next frame $t+1$, the NSD is calculated by the black box at the BS and a reward signal is obtained. The BS broadcasts this reward signal to each agent and a new state is obtained for each agent. Each agent acts accordingly in the next time step. The reward received by each agent is common to incite a cooperative behavior among devices. Thanks to the simplicity of the online competitive algorithm, our approach solves perfectly the curse of dimensionality issue in reinforcement learning. The details of the learning process are given in the sequel.

We model GPA with multiple frames as an online deterministic Markov decision process (MDP). This modelling is important to apply reinforcement learning technique and helped us to transform the problem into an online (stochastic) shortest path problem. The MDP is deterministic because the transition probabilities are known. The corresponding transition graph (TG), that models the state space, the action space, the transition function, and the reward function, is constructed as follows. An example of this TG is given in Fig.~\ref{fig:dag}.
\begin{itemize}
\item \textbf{State}: a state (or a node) in the TG is a tuple $(\mathbf{e}^t,t)$, for $t=2,\ldots,k+1$, where $\mathbf{e}^t=[e_{1}^t,e_{2}^t,\ldots,e_{m}^t]^\top$ represents the remaining battery level of the devices in frame $t$. When $t=1$, the node $\mathbf{s}\coloneq(\mathbf{e}^1,1)$ is called the starting node, where $e_{i}^1=\upper{e}_i$ for all $i$. There is a terminal node denoted by $\mathbf{t}\coloneq(\mathbf{e}^{k+2},k+2)$, where $e_{i}^{k+2}=0$ for all $i$.

\item \textbf{Transition}: for $t=1,2,\ldots,k+1$, a transition from $(\mathbf{e}^t_1,t)$ to $(\mathbf{e}^{t+1}_2,t+1)$ happens with probability one if and only if $\mathbf{e}^t_1-\mathbf{e}^{t+1}_2\succcurlyeq\mathbf{0}$. No other transition is allowed. This means that if the device has some remaining power in its battery, then it can transition to a new state, otherwise it cannot.

\item \textbf{Action}: for $t=1,2,\ldots,k+2$, the action set corresponding to state $(\mathbf{e}^t,t)$ is given by the Cartesian product $[e_{1}^t]_{\tau_1}\times[e_{2}^t]_{\tau_2}\times\cdots\times[e_{m}^t]_{\tau_m}$, that is, an action taken in state $(\mathbf{e}_1^t,t)$ and transitions to state $(\mathbf{e}_2^{t+1},t+1)$ is a transmission power vector $\mathbf{p}^t=[p_{1}^t,p_{2}^t,\ldots,p_{m}^t]^\top$. In other words, the possible actions in state $(\mathbf{e}_1^t,t)$ are given by the outgoing edges of node $(\mathbf{e}_1^t,t)$. Denote by $m_i\coloneq|[\upper{e}_i]_{\tau_i}|=\tau_i+1$ the power level of device $i$ and by $m_x\coloneq\Pi_{i=1}^mm_i$. The TG contains $2+km_x$ states and $m_x(m_x(k-1)+k+3)/2$ directed edges.

\item \textbf{Reward}: the reward of choosing action $\mathbf{p}^t=[p_{1}^t,p_{2}^t,\ldots,p_{m}^t]^\top$ in state $(\mathbf{e}^t,t)$ is given by the NSD in frame $t$, which can be obtained by applying the previously proposed online competitive algorithm, \textsc{bms}.
\end{itemize}
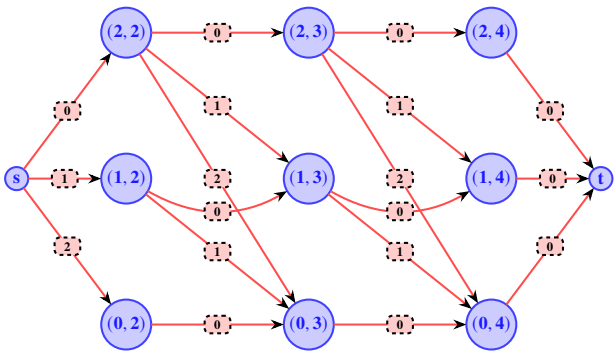
\begin{figure}[ht!]
  \centering
  \resizebox{.45\textwidth}{!}{%
  \begin{tikzpicture}%
  \Vertex[Math,L={\color{blue!90}\mathbf{s}},x=2,y=-5]{A}
  \Vertex[Math,L={\color{blue!90}\mathbf{(2,2)}},x=5,y=-1]{B}
  \Vertex[Math,L={\color{blue!90}\mathbf{(1,2)}},x=5,y=-5]{C}
  \Vertex[Math,L={\color{blue!90}\mathbf{(0,2)}},x=5,y=-9]{D}
  \Vertex[Math,L={\color{blue!90}\mathbf{(2,3)}},x=10,y=-1]{E}
  \Vertex[Math,L={\color{blue!90}\mathbf{(1,3)}},x=10,y=-5]{F}
  \Vertex[Math,L={\color{blue!90}\mathbf{(0,3)}},x=10,y=-9]{G}
  \Vertex[Math,L={\color{blue!90}\mathbf{(2,4)}},x=15,y=-1]{H}
  \Vertex[Math,L={\color{blue!90}\mathbf{(1,4)}},x=15,y=-5]{I}
  \Vertex[Math,L={\color{blue!90}\mathbf{(0,4)}},x=15,y=-9]{J}
  \Vertex[Math,L={\color{blue!90}\mathbf{t}},x=18,y=-5]{K}
  \Edge[style={->,thick},label = 0](A)(B)
  \Edge[style={->,thick},label = 1](A)(C)
  \Edge[style={->,thick},label = 2](A)(D)
  \Edge[style={->,thick},label = 0](B)(E)
  \Edge[style={->,thick},label = 1](B)(F)
  \Edge[style={->,thick},label = 2](B)(G)
  \Edge[style={->,thick,bend right},label = 0](C)(F)
  \Edge[style={->,thick},label = 1](C)(G)
  \Edge[style={->,thick},label = 0](D)(G)
  \Edge[style={->,thick},label = 0](E)(H)
  \Edge[style={->,thick},label = 1](E)(I)
  \Edge[style={->,thick},label = 2](E)(J)
  \Edge[style={->,thick,bend right},label = 0](F)(I)
  \Edge[style={->,thick},label = 1](F)(J)
  \Edge[style={->,thick},label = 0](G)(J)
  \Edge[style={->,thick},label = 0](H)(K)
  \Edge[style={->,thick},label = 0](I)(K)
  \Edge[style={->,thick},label = 0](J)(K)
  \end{tikzpicture}
  }
  \caption{An instance of the TG of one device with $[2]_2=\{0,1,2\}$ and three frames. The rounded squares in the middle of the edges represent the actions.}
  \label{fig:dag}
\end{figure}

Under this modelling, GPA can be seen as an $\mathbf{s}$-$\mathbf{t}$ shortest path problem in the corresponding TG, or equivalently, as finding the $\mathbf{s}$-$\mathbf{t}$ path with the highest reward (by transforming rewards to losses we can move from shortest path to longest path). Nonetheless, finding such an $\mathbf{s}$-$\mathbf{t}$ path is too complex because the number of nodes and the number of edges in the TG is exponentially large, e.g., for 3 frames, 15 devices, and a power level of two ($m_i=2$ for all $i$), the TG contains approximately 100000 nodes and $10^9$ edges. 

Due to the curse of dimensionality, we follow a \textit{distributed multi-agent} approach to solve the power allocation learning problem. The approach is distributed in the sense that each device learns its own $\mathbf{s}$-$\mathbf{t}$ shortest path by communicating with the BS. The distributed multi-agent approach uses the idea of agent cooperation through proper reward design. The reward is designed to be common for all devices to incite them to act cooperatively. This reward is in fact returned by the previously proposed online competitive algorithm, \textsc{bms}, in each frame. It is equal to the NSD in each frame. Each device uses a modified version of \textsc{exp3}~\cite{10.1137/S0097539701398375}---a popular reinforcement learning algorithm for the adversarial multi-armed bandit problem. For comparison purposes, we also adopt the classical tabular Q-learning algorithm~\cite{10.5555/551283}.

\begin{remark}[Notes on \textsc{exp3} and deep reinforcement learning]\label{remark:dql}
    \textsc{exp3} is generally used to solve non-stochastic learning problem, i.e., when the input to the learning problem is non-stochastic and thus it varies arbitrarily without explicitly known distribution. Specifically, \textsc{exp3} is proposed to solve the non-stochastic multi-armed bandit problem~\cite{10.1137/S0097539701398375}, contrary for example to the well-known upper confidence bound (\textsc{ucb}) algorithm that is proposed to solve the stochastic multi-armed bandit problem. The proposed framework---combining \textsc{exp3} and online competitive algorithms---is shown to be simple (in terms of complexity), scalable, and robust.

    Applying advanced reinforcement learning algorithms, such as deep reinforcement learning to solve GPA is very appealing. However, it is also challenging due to the intensive computation training tasks that should be carried out either by the BS or by each device. In both cases, this might not be a scalable approach in a massive access IoT network with hundreds of devices and thus the deep neural network would have thousands of input elements. It might be feasible to perform such intensive-training tasks but it will surely require more computing capabilities as well as more complexity and running times than simple reinforcement learning frameworks. Further, the training should be re-executed once every time a major change happens in the IoT environment. Contrary, the proposed learning framework is very simple and it only requires a few tens of iterations in which simple probability updates are performed. Despite this, the investigation of deep reinforcement learning to solve GPA will be studied for our future works.
\end{remark}

\subsection{\textsc{exp3}-based Distributed Learning}
\textsc{exp3} is based on exponential-weighting for exploration and exploitation and is proposed to solve the non-stochastic (adversarial) multi-armed bandit problem~\cite{10.1137/S0097539701398375}. Each device learns its own $\mathbf{s}$-$\mathbf{t}$ path by applying a modified version of \textsc{exp3}. Each device has its own TG. As before, in the distributed implementation, the state space, the action space, the transition function, and the reward function are constructed as follows. See Fig.~\ref{fig:dag} for an example.
\begin{itemize}
\item \textbf{State}: a state in each TG $i$ is given by $(e_{i}^t,t)$ where $e_i^t$ represents the remaining energy level at frame $t$ in device $i$'s battery with $e_{i}^1=\upper{e}_i$.

\item \textbf{Transition}: for $t=1,2,\ldots,k+1$, a transition from $(e_{i1}^t,t)$ to $(e_{i2}^{t+1},t+1)$ happens with probability one if and only if $e_{i1}^t-e_{i2}^{t+1}\geqslant0$. No other transition is allowed. This means that if the device has some remaining power in its battery, then it can transition to a new state, otherwise it cannot.

\item \textbf{Action}: for any state $(e_{i}^t,t)$ of TG $i$, an action is given by the transmission power $p_{i}^t\in[e_{i}^t]_{\tau_i}$.

\item \textbf{Reward}: normally, when device $i$, in state $(e_{i}^t,t)$, chooses action $p_{i}^t\in[e_{i}^t]_{\tau_i}$, its reward is a binary number that represents whether or not it is served. Designing the rewards in this way teaches the devices to act selfishly and thus does not necessarily give a good outcome, i.e., the total NSD could be very low because each one will learn to use its transmission power to get served regardless of others (see remark~\ref{selfish}). It is thus necessary to redesign the rewards to improve the learning outcome. Instead of the binary rewards, each device receives its reward as the NSD in each frame. This can be acquired by information feedback between the devices and the BS.
\end{itemize}

The main lines of the learning algorithm, called the \textsc{pl} algorithm for \textit{path-learning} is given for each round as follows.
\begin{itemize}
    \item Device $i$ chooses an action $p_{i}^t$ for each frame $t$ according to some probability, i.e., it chooses an $\mathbf{s}_i$-$\mathbf{t}_i$ path in TG $i$. We denote this path by the vector $\mathbf{p}_i=[p_{i}^1,p_{i}^2,\ldots,p_{i}^k]^\top$.
    \item Device $i$ sends its chosen $\mathbf{s}_i$-$\mathbf{t}_i$ path to the BS.
    \item The BS runs the online per-frame competitive algorithm, \textsc{bms}, at frame $t$ with power allocation $\mathbf{p}^t=[p_{1}^t,p_{2}^t,\ldots,p_{m}^t]^\top$ and calculates the NSD.
    \item The BS broadcasts the rewards to each device (the reward received by device $i$ is the NSD in frame $t$). That is, device $i$  knows, not only the rewards of its chosen $\mathbf{s}_i$-$\mathbf{t}_i$ path, but also the rewards in each edge of that path.
    \item Device $i$ updates the probabilities.
\end{itemize}

\textsc{pl} operates in rounds where, in each round, it is applied at device $i$ that chooses an $\mathbf{s}_i$-$\mathbf{t}_i$ path according to some probability (proportional to the path weight). This probability is chosen to follow a distribution over the set of all $\mathbf{s}_i$-$\mathbf{t}_i$ paths in order to get a mixture between exponential weighting of biased estimates of the rewards and uniform distribution to ensure a sufficiently large exploration of each edge of any $\mathbf{s}_i$-$\mathbf{t}_i$ path. After choosing an $\mathbf{s}_i$-$\mathbf{t}_i$ path, device $i$ gets to know the rewards on each edge of that path, i.e, it gets to know the NSD in each frame. Then, \textsc{pl} updates the probability distribution (by updating the paths weights) and continues similarly. 

We notice that every TG $i$ has $2+km_i$ nodes and $m_i(m_i(k-1)+k+3)/2$ directed edges. Every path in TG $i$ has length $k+1$. Let $\mathbb{P}_i$ be the set of all $\mathbf{s}_i$-$\mathbf{t}_i$ paths in TG $i$ and let $\sigma_i\coloneq|\mathbb{P}_i|$ denotes the number of such paths. We can prove that $\sigma_i=\binom{k+m_i-1}{k}$, which is exponentially large and thus choosing the paths in this way according to their weights is not efficient. However, a simple modification can improve the algorithm enormously~\cite{1633787}. First, instead of assigning weights to paths, they are assigned to edges. Second, we construct a set of edge-covering $\mathbf{s}_i$-$\mathbf{t}_i$ paths $\mathbb{C}_i$, which is defined as the set of paths in TG $i$ such that for any edge $e$ in TG $i$, there is a path $\mathbf{p}_i$ in $\mathbb{C}_i$ such that $e\in\mathbf{p}_i$. Such an edge-covering paths $\mathbb{C}_i$ can be obtained in $\mathscr{O}(km_i^2+km_i\lg(km_i))$ time using Dijkstra's algorithm where $|\mathbb{C}_i|=\mathscr{O}(km_i^2)$. Now, instead of each path, each edge $e$ of TG $i$ is assigned a weight $w(e)$ (initialized to one for each edge at the beginning of the rounds) and the weight of an $\mathbf{s}_i$-$\mathbf{t}_i$ path is given by the product of the weights of its edges. For each round, \textsc{pl}, applied at device $i$, chooses an $\mathbf{s}_i$-$\mathbf{t}_i$ path (1) uniformly from $\mathbb{C}_i$ with probability $\gamma$ or (2) according to the paths weights with probability $1-\gamma$. If the latter is to be done, then the $\mathbf{s}_i$-$\mathbf{t}_i$ path can be chosen by adding its vertexes one-by-one according to edges' weights (and not to paths' weights)~\cite{1633787}. Next, \textsc{pl} finds the probability of choosing each edge in the TG $i$, which can also be done using edges' weights only. (It can be proven that choosing paths and updating the edges' probabilities can be done efficiently based on paths kernels and dynamic programming~\cite{10.1007/3-540-45435-7_6}.) Then, for each frame (or equivalently for each edge), the rewards are obtained using \textsc{bms}, where the reward at any edge $r$ is normalized by the probability of that edge $q(e)$, i.e., the normalized reward is $(\beta+r\mathds{1}_{\{e\in\mathbf{p}_i\}})/q(e)$, with $\mathds{1}_{\mathbb{A}}$ denotes the indicator function and $\beta\in(0,1]$. Finally, the edges' weights are updated as $w(e)\gets w(e)e^{\eta r}$ where $\eta>0$.

The per-round complexity of \textsc{pl} is given by $\mathscr{O}(kmm_i^2+knm\lg m)$, where $\mathscr{O}(knm\lg m)$ is the complexity of applying \textsc{bms} in all frames and $\mathscr{O}(kmm_i^2)$ is the complexity of choosing the paths according to the edges' weights and updating the probability of each edge.

\subsection{Tabular-based Distributed Learning}
We use the tabular Q-learning algorithm~\cite{10.5555/551283}. The Q-learning algorithm is called \textsc{ql} and it proceeds in episodes. In each episode, each device $i$ chooses an $\mathbf{s}_i$-$\mathbf{t}_i$ path, receives a reward, and updates its Q-table. Precisely, each device $i$ has a Q-table $Q_i(s,a)$ that measures the quality of a state-action combination $(s,a)$, where $s$ represents a node in the TG $i$ and $a$ represents a chosen transmission power in state $s$. For each episode, each device $i$ starts the learning in the initial state $\mathbf{s}_i$. For each step in that episode, that is for each frame $t$, device $i$ chooses a possible action (according to its state) using the epsilon-greedy approach and moves to the next state $s'$. Once all devices choose their actions, the BS runs the online competitive algorithm, \textsc{bms}, in frame $t$ and fed back the rewards to each device (device $i$ receives the NSD in frame $t$). Next, each device moves to the next state and updates its Q-table. As soon as the last frame is reached and the Q-table is updated, the devices move to the next episode and the Q-learning algorithm continues. Updating the Q-table is done as follows: $Q(s,a)\gets Q(s,a)+\alpha(r+\max_{a}Q(s',a)-Q(s,a))$. The per-episode complexity of \textsc{ql} is given by $\mathscr{O}(knm\lg m)$, where $\mathscr{O}(nm\lg m)$ is the complexity of applying \textsc{bms} in each frame and updating the Q-table.

\section{Simulation Results}\label{sec:simulations}
This section illustrates the performance of the proposed algorithms through computer simulations. We consider a geographical zone modelled by a square of side $1000$ meters. The BS is located at the centre of this zone whereas the devices are randomly and uniformly distributed inside the square. The simulations parameters are based on 3GPP specifications~\cite[p.~481]{3gpp.45.820} as in~\cite{7165674,8764608}. The carrier frequency is $f_c=900$ MHz and the path-loss (in dB) at $f_c$ is given by $120.9+37.6\log(\text{dist}_{i}^t)+\alpha_{\text{G}}+\alpha_{\text{L}}$~\cite[p.~481]{3gpp.45.820}, where $\text{dist}_{i}^t$ is the distance (in km) between device $i$ and the BS at frame $t$, $\alpha_{\text{G}}=-4$ dB represents the antenna gain and $\alpha_{\text{L}}=10$ dB is the penetration loss. Flat Rayleigh fading is also considered and thus $g_{ij}^t$ includes the previous path-loss model as well as an exponential random variable with unit parameter. The power spectral density of the noise is $-174$ dBm/Hz and the noise figure is $5$ dB. Unless specified otherwise, the next parameters are fixed as follows. Each device $i$ has a maximum transmission power of $\upper{e}_i=23$ dBm~\cite[p.~481]{3gpp.45.820}. The group size is $M=2$. The bandwidth is $200$ kHz and the bandwidth of a single RB is $200/n$ kHz where $n$ is the total number of RBs. The data requirements of the devices follow a uniform distribution as $L_i^t\sim\mathrm{unif}\{0, L_{\text{max}}\}$ with $L_{\text{max}}=100$ kbits. The arrival times are given by $a_i^t\sim\mathrm{unif}\{1,n\}$ and the deadlines are given by $d_i^t\sim\mathrm{unif}\{a_i^t+1,n+1\}$. The optimal offline algorithm to solve~\eqref{pb:1} in the single frame model, denoted \textsc{opt}, is based on AMPL modelling~\cite{10.5555/2804729.2804730} and using the CPLEX solver.

The next figures show the results for the single frame model in which the problem involves only NOMA grouping.
\begin{figure}[htpb!]
\centering
\captionsetup{justification=centering,margin=1cm}
	\includegraphics[scale=0.2]{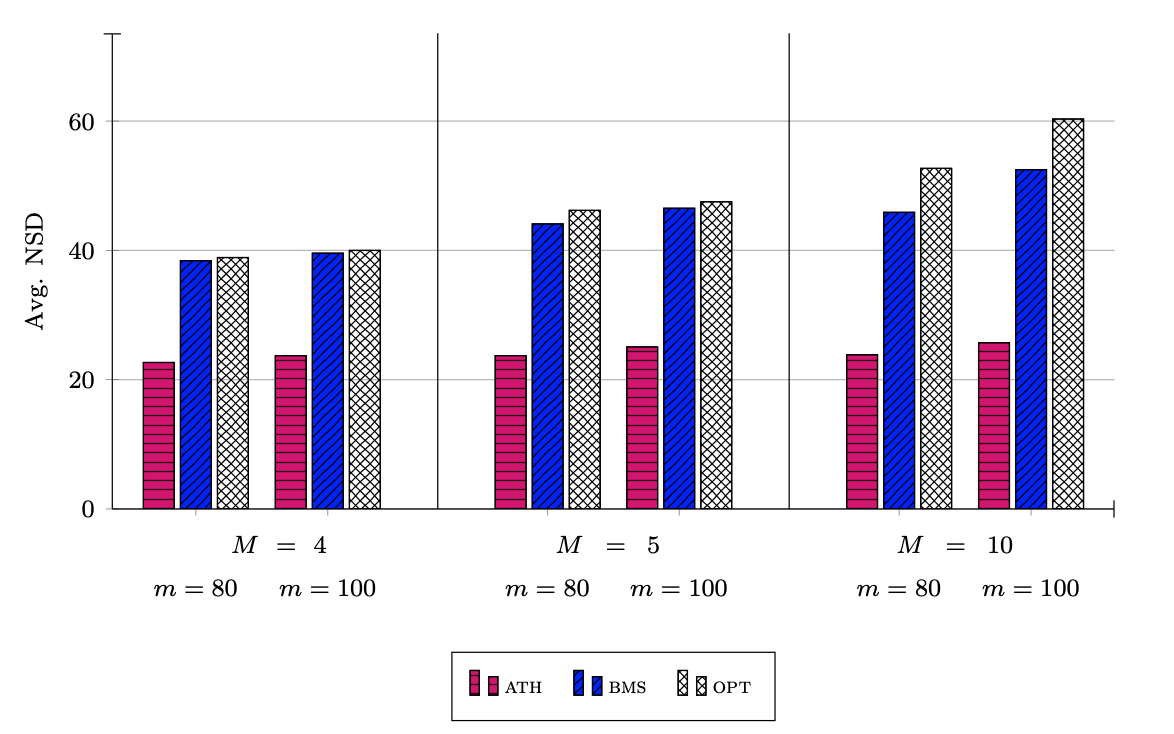}
  \caption{Impact of $M$ for the single frame model.}
  \label{fig:M}
\end{figure}%
Fig.~\ref{fig:M} shows the impact of $M\geqslant2$ on \textsc{opt}, \textsc{ath} and \textsc{bms}\footnote{As disccussed previously, \textsc{zz} is proposed only for $M=2$ and thus it is not included in the comparison in Fig.~\ref{fig:M}.}. We can see that as $M$ increases, the NSD increases for all algorithms. Nonetheless, the NSD increases slightly in \textsc{ath} because the number of clusters is inversely proportional to $M$ as described in~\cite{7557079}, i.e., the number of clusters is equal to the number of devices divided by $M$. The proposed online competitive algorithm, \textsc{bms}, outperforms always the benchmark \textsc{ath} and is close-to-optimal. We can further notice that the gap between \textsc{opt} and \textsc{bms} is much less than the theoretical performance guarantee of 50\% proven in Theorems~\ref{the2com} and~\ref{theo:mub}.

 \begin{table*}[b!]
  \centering
  \setlength\tabcolsep{0pt}
  \caption{Running times (in seconds) for the algorithms of the single frame model.}
  \label{tab:runtime}
  \begin{tabular}{|M{3cm}|M{4.5cm}||M{4.5cm}||M{4.5cm}|}\hline
    Algorithms   & Configuration 1 & Configuration 2 & Configuration 3\\\hline
    \textsc{bms} & 1.27e-4         & 3.03e-4         & 7.77e-4\\\hline
    \textsc{ath} & 40.90e-4        & 75.92e-4        & 102.53e-4\\\hline
    \textsc{zz}  & 190.46e-4       & 936.37e-4       & NaN\\\hline
    \textsc{opt} & 19050e-4        & 48837.6e-4      & 1499630e-4\\\hline
  \end{tabular}
\end{table*}

\begin{figure}[ht!]
\centering
\captionsetup{justification=centering,margin=1cm}
  \resizebox{.44\textwidth}{!}{%
    \begin{tikzpicture}
    \tikzset{every pin/.style={draw=black,fill=yellow!20,rectangle,rounded corners=3pt,font=\scriptsize},}
    \begin{axis}[
    xlabel={Number of devices ($m$)},
    ylabel={Avg. NSD},
    set layers,
    grid=both,
    legend cell align=left,
    xmin=20,
    xmax=100,
    ymin=10,
    ymax=61,
    xtick={20,40,...,100},
    x label style={font=\footnotesize},
    y label style={font=\footnotesize}, 
    ticklabel style={font=\footnotesize},
    legend style={at={(.22,.85)},anchor=east,font=\scriptsize},
    ]
    
    \addplot[dashed,color=black,mark=asterisk,mark options={scale=1,solid}] coordinates {
        (20, 17.254) (40, 19.82) (60, 19.982) (80, 19.996) (100, 20)
    };\addlegendentry{\textsc{opt}}
    \addplot[dashed,color=blue,mark=o,mark options={scale=1,solid}] coordinates {
        (20, 16.206) (40, 19.7) (60, 19.942) (80, 19.986) (100, 19.998)
    };\addlegendentry{\textsc{bms}}
        \addplot[dashed,color=red,mark=triangle,mark options={scale=1,solid}] coordinates {
            (20, 16.288) (40, 18.61) (60, 19.24) (80, 19.47) (100, 19.658)
        };\addlegendentry{\textsc{zz}}
    \addplot[dashed,color=magenta,mark=halfsquare right*,mark options={scale=1,solid}] coordinates {
        (20, 12.414) (40, 15.974) (60, 17.648) (80, 18.494) (100, 18.896)
    };\addlegendentry{\textsc{ath}}
    
    \addplot[dashed,color=black,mark=asterisk,mark options={scale=1,solid}] coordinates {
        (20, 19.464) (40, 34.664) (60, 38.134) (80, 38.826) (100, 39.510)
    };
    \addplot[dashed,color=blue,mark=o,mark options={scale=1,solid}] coordinates {
        (20, 18.138) (40, 30.308) (60, 36.564) (80, 38.762) (100, 39.508)
    };
    \addplot[dashed,color=red,mark=triangle,mark options={scale=1,solid}] coordinates {
    	(20, 18.814) (40, 30.066) (60, 33.94) (80, 35.926) (100, 37.004)
	};
    \addplot[dashed,color=magenta,mark=halfsquare right*,mark options={scale=1,solid}] coordinates {
        (20, 16.636) (40, 22.968) (60, 26.008) (80, 28.542) (100, 30.694)
    };        
    \addplot[dashed,color=black,mark=asterisk,mark options={scale=1,solid}] coordinates {
        (20, 18.84) (40, 33.55) (60, 44.39) (80, 52.69) (100, 60.2)
    };
    \addplot[dashed,color=blue,mark=o,mark options={scale=1,solid}] coordinates {
        (20, 17.506) (40, 29.462) (60, 38.552) (80, 46.012) (100, 52.592)
    };
    
    \node[coordinate] (A) at (axis cs:80,49.1) {}; 
    \node[coordinate,pin={[align=left,pin distance=.5cm,pin edge={black,thick}]130:{$n=10,M=10$}}] at (axis cs:78.4,50) {}; 
    \node[coordinate] (B) at (axis cs:80,33.8) {}; 
    \node[coordinate,pin={[align=left,pin distance=.4cm,pin edge={black,thick},reset label anchor]270:{$n=20,M=2$}}] at (axis cs:80,27.6) {}; 
    \node[coordinate] (C) at (axis cs:40,18) {}; 
    \node[coordinate,pin={[align=left,pin distance=.5cm,pin edge={black,thick}]315:{$n=10,M=2$}}] at (axis cs:41.5,18.95) {}; 
    \end{axis}
    \draw[black,thick] (A) ellipse (0.15 and 0.5); 
    \draw[black,thick] (B) ellipse (0.15 and 0.7); 
    \draw[black,thick] (C) ellipse (0.15 and 0.35); 
    \end{tikzpicture}
  }
  \caption{Impact of $m$ for the single frame model.}
  \label{fig:1}
\end{figure}%
Fig.~\ref{fig:1} shows the performance of the proposed online competitive algorithm, \textsc{bms}, for different values of $M$ and $n$ against \textsc{ath}, \textsc{zz}, and \textsc{opt}. When the number of RBs $n$ or the group size $M$ increases, the number of served devices increases faster with $m$. When the number of devices is small and the number of RBs is large, the offline algorithm \textsc{zz} achieves slightly better performance compared to our proposed algorithm \textsc{bms}, despite being online. This is might be due to the heuristic approach used in \textsc{zz} to find a maximal independent set in the graph $H$ as discussed in~\ref{SubSec1}. Another point is that \textsc{bms} achieves much better performance compared to \textsc{ath}, because (1) the latter mainly optimizes the sum-rate objective and not the NSD and further (2) the pairs of devices in the latter are fixed a priori (thus, because the stringent constraints in GPA very few pairs can satisfy these constraints according to NOMA). Lastly, despite the online nature of \textsc{bms} and its $50\%$ theoretical worst-case gap, Fig.~\ref{fig:1} shows that its performance is not very far from that of \textsc{opt} even for large values of $n$ and $M$. Indeed, the largest gap between \textsc{bms} and \textsc{opt} as shown in Fig.~\ref{fig:1} is about $87\%$ which is much better than the $50\%$ theoretical gap proven in Theorems~\ref{the2com} and~\ref{theo:mub}.

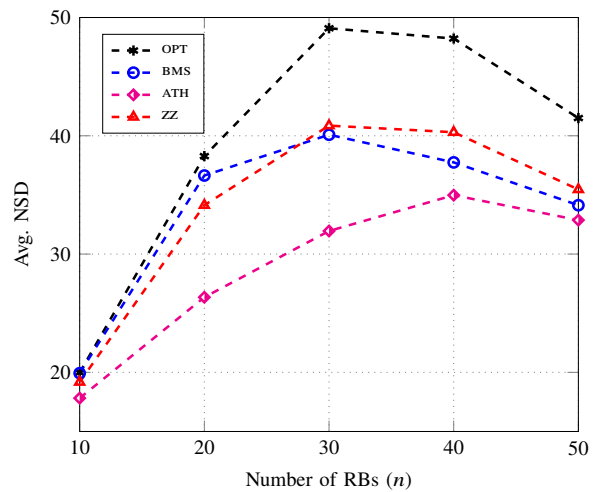
\begin{figure}[ht!]
  \centering
  \captionsetup{justification=centering,margin=.7cm}
  \resizebox{.44\textwidth}{!}{%
    \begin{tikzpicture}
    \tikzset{every pin/.style={draw=black,fill=yellow!20,rectangle,rounded corners=3pt,font=\scriptsize},}
    \begin{axis}[
    xlabel={Number of RBs ($n$)},
    ylabel={Avg. NSD},
    set layers,
    grid=both,
    legend cell align=left,
    xmin=10,
    xmax=50,
    ymin=15,
    ymax=50,
    xtick={10,20,...,50},
    x label style={font=\footnotesize},
    y label style={font=\footnotesize}, 
    ticklabel style={font=\footnotesize},
    legend style={at={(.25,.84)},anchor=east,font=\scriptsize},
    ]
    \addplot[dashed,color=black,mark=asterisk,mark options={scale=1,solid}] coordinates {
        (10, 19.98) (20, 38.28) (30, 49.09) (40, 48.22) (50, 41.5)
    };\addlegendentry{\textsc{opt}}
    \addplot[dashed,color=blue,mark=o,mark options={scale=1,solid}] coordinates {
            (10, 19.93) (20, 36.64) (30, 40.09) (40, 37.75) (50, 34.13)
        };\addlegendentry{\textsc{bms}}
    \addplot[dashed,color=magenta,mark=halfsquare right*,mark options={scale=1,solid}] coordinates {
            (10, 17.82) (20, 26.35) (30, 31.95) (40, 34.98) (50, 32.87)
    };\addlegendentry{\textsc{ath}}
    \addplot[dashed,color=red,mark=triangle,mark options={scale=1,solid}] coordinates {
            (10, 19.18) (20, 34.13) (30, 40.86) (40, 40.3) (50, 35.46)
	};\addlegendentry{\textsc{zz}}        
    \end{axis}
    \end{tikzpicture}
  }
  \caption{Impact of $n$ for the single frame model ($M=2$).}
  \label{fig:2}
  \end{figure}
Fig.~\ref{fig:2} presents the performance of \textsc{bms} for $m=60$ and $M=2$ against \textsc{ath}, \textsc{zz} and \textsc{opt}. When $m$ and $M$ are fixed, there is an optimal value of $n$ at which the performance is maximized. When $n$ increases above this optimal value, the NSD starts to decrease since the bandwidth of each RB becomes small. In other words, when $n$ continues to grow which decreases the bandwidth of each RB, the interference inside each NOMA group becomes intolerable and the devices cannot meet their requirements. Lately, the performance of \textsc{bms} is still the best amongst the non-optimal algorithms except for large $n$ where \textsc{zz} becomes better (since it is offline and it explores more nodes in $H$) at the expense of higher running time complexity and more powerful capabilities of seeing future inputs. Finally, \textsc{bms} achieves close-to-optimal performance for different $n$ and the worst gap is about $78\%$ which is much better than the proven $50\%$ theoretical gap (see Theorems~\ref{the2com} and~\ref{theo:mub}). 

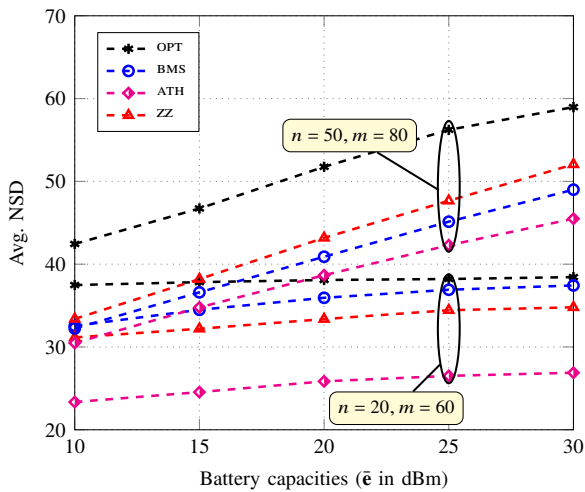
\begin{figure}[ht!]
  \centering
  \captionsetup{justification=centering,margin=1cm}
  \resizebox{.44\textwidth}{!}{%
    \begin{tikzpicture}
    \tikzset{every pin/.style={draw=black,fill=yellow!20,rectangle,rounded corners=3pt,font=\scriptsize},}
    \begin{axis}[
    xlabel={Battery capacities ($\upper{\mathbf{e}}$ in dBm)},
    ylabel={Avg. NSD},
    set layers,
    grid=both,
    legend cell align=left,
    xmin=10,
    xmax=30,
    ymin=20,
    ymax=70,
    xtick={10,15,...,30},
    x label style={font=\footnotesize},
    y label style={font=\footnotesize}, 
    ticklabel style={font=\footnotesize},
    legend style={at={(.25,.84)},anchor=east,font=\scriptsize},
    ]
    \addplot[dashed,color=black,mark=asterisk,mark options={scale=1,solid}] coordinates {
        (10, 37.48) (15, 37.83) (20, 38.09) (25, 38.21) (30, 38.44)
    };\addlegendentry{\textsc{opt}}
    \addplot[dashed,color=blue,mark=o,mark options={scale=1,solid}] coordinates {
            (10, 32.55) (15, 34.49) (20, 35.94) (25, 36.91) (30, 37.43)
    };\addlegendentry{\textsc{bms}}
    \addplot[dashed,color=magenta,mark=halfsquare right*,mark options={scale=1,solid}] coordinates {
            (10, 23.35) (15, 24.55) (20, 25.85) (25, 26.51) (30, 26.89)
    };\addlegendentry{\textsc{ath}}
    \addplot[dashed,color=red,mark=triangle,mark options={scale=1,solid}] coordinates {
            (10, 31.15) (15, 32.2) (20, 33.37) (25, 34.44) (30, 34.79)
        };\addlegendentry{\textsc{zz}}
        
    \addplot[dashed,color=black,mark=asterisk,mark options={scale=1,solid}] coordinates {
        (10, 42.44) (15, 46.73) (20, 51.76) (25, 56.23) (30, 58.97)
    };
    \addplot[dashed,color=blue,mark=o,mark options={scale=1,solid}] coordinates {
            (10, 32.24) (15, 36.6) (20, 40.9) (25, 45.13) (30, 49.0)
    };
    \addplot[dashed,color=magenta,mark=halfsquare right*,mark options={scale=1,solid}] coordinates {
            (10, 30.48) (15, 34.78) (20, 38.67) (25, 42.29) (30, 45.49)
    };
    \addplot[dashed,color=red,mark=triangle,mark options={scale=1,solid}] coordinates {
            (10, 33.39) (15, 38.2) (20, 43.18) (25, 47.66) (30, 52.05)
        };
        
    \node[coordinate] (A) at (axis cs:25,49.4) {}; 
    \node[coordinate,pin={[align=left,pin distance=.6cm,pin edge={black,thick}]120:{$n=50,m=80$}}] at (axis cs:24.6,48.95) {}; 
    \node[coordinate] (B) at (axis cs:25,32.2) {}; 
    \node[coordinate,pin={[align=left,pin distance=.85cm,pin edge={black,thick},reset label anchor]225:{$n=20,m=60$}}] at (axis cs:24.65,27.88) {}; 
    \end{axis}
    \draw[black,thick] (A) ellipse (0.15 and 0.9); 
    \draw[black,thick] (B) ellipse (0.15 and 0.75); 
    \end{tikzpicture}
  }
  \caption{Impact of $\upper{\mathbf{e}}$ for the single frame model.}
  \label{fig:3}
\end{figure}
Fig.~\ref{fig:3} presents the performance of \textsc{bms} against \textsc{ath}, \textsc{zz}, and \textsc{opt} when the battery capacity changes. Increasing the battery capacity can increase the NSD quickly (the increase rate is faster when the number of RBs is larger). However, since the number of RBs and the group size are limited, the increase rate slows down as the battery capacity increases, and thus the curves start to converge. 
Despite being online and much simpler, \textsc{bms} is very close to \textsc{opt}. As for \textsc{zz}, since it is offline, it outperforms \textsc{bms} for large $n$ but it is much complex. 

In table~\ref{tab:runtime}, we measure the running time of the different algorithms. These measurements are performed using a desktop computer with Linux operating system, architecture x86\_64, central processing unit (CPU) on mode 64-bit, 8 CPUs cores, Intel(R) Core(TM) i7-9700K CPU @ 3.60GHz, maximum CPU frequency 4900 MHz, and minimum CPU frequency  800 MHz. The module \texttt{timeit} is used in Python for all algorithms except for \textsc{opt} in which the command \texttt{\_total\_solve\_time} is used in AMPL. The values presented in table~\ref{tab:runtime} are averaged over 5000 calculations. The columns of the table (from left to right) correspond to the following configurations. Configuration 1 is when $n=10$, $m=10$, and $M=2$. Configuration 2 is when $n=10$, $m=20$, and $M=2$. Configuration 3 is when $n=10$, $m=40$, and $M=10$. We can see that \textsc{bms} achieves the lowest running time regardless of the configuration. On the other hand, \textsc{opt} has the highest running time and as the number of devices or the group size increases, the running time exponentially increases. As for \textsc{ath} and \textsc{zz}, we can see that the latter has the worst running time since it requires generating all devices' pairs. Lastly, when $M\ne2$, we add ``Not a Number'' for \textsc{zz} in the third row, last column to indicate that \textsc{zz} is not defined in that case.

\begin{figure}[ht!]
  \centering
  \captionsetup{justification=centering,margin=1cm}
  \resizebox{.44\textwidth}{!}{%
    \begin{tikzpicture}
    \tikzset{every pin/.style={draw=black,fill=yellow!20,rectangle,rounded corners=3pt,font=\scriptsize},}
    \begin{axis}[
    xlabel={Number of devices ($m$)},
    ylabel={Avg. NSD},
    set layers,
    grid=both,
    legend cell align=left,
    xmin=1000,
    xmax=2000,
    ymin=10,
    ymax=300,
    xtick={1000,1200,1400,...,2000},
    xticklabels={1000,1200,1400,1600,1800,2000},
    x label style={font=\footnotesize},
    y label style={font=\footnotesize}, 
    ticklabel style={font=\footnotesize},
    legend style={at={(1,.22)},anchor=east,font=\scriptsize},
    ]
    \addplot[color=red,mark=o,mark options={scale=1,solid}] coordinates {
            (1000, 193.42) (1100, 203.03) (1200, 212.54) (1300, 221.72) (1400, 231.41) (1500, 239.28) (1600, 248.27) (1700, 256.59) (1800, 264.93) (1900, 271.92) (2000, 279.82)
    };\addlegendentry{\textsc{bms} ($n=M=20, L_{\text{max}}=100$ kbits)}
    \addplot[dashed,color=blue,mark=diamond,mark options={scale=1,solid}] coordinates {
            (1000, 185.73) (1100, 191.61) (1200, 195.46) (1300, 197.64) (1400, 198.88) (1500, 199.15) (1600, 199.35) (1700, 199.74) (1800, 199.9) (1900, 199.79) (2000, 199.92)
    };\addlegendentry{\textsc{bms} ($n=10, M=20, L_{\text{max}}=100$ kbits)}
    \addplot[dotted,color=black,mark=asterisk,mark options={scale=1,solid}] coordinates {
            (1000, 196.86) (1100, 206.95) (1200, 216.0) (1300, 225.79) (1400, 235.21) (1500, 244.28) (1600, 251.63) (1700, 260.46) (1800, 270.18) (1900, 277.06) (2000, 285.05)
    };\addlegendentry{\textsc{bms} ($n=M=40, L_{\text{max}}=100$ kbits)}

    \addplot[densely dashed,color=magenta,mark=star,mark options={scale=1,solid}] coordinates {
            (1000, 133.2) (1100, 139.89) (1200, 146.77) (1300, 153.6) (1400, 159.48) (1500, 165.17) (1600, 170.87) (1700, 176.79) (1800, 182.51) (1900, 188.01) (2000, 193.3)
    };\addlegendentry{\textsc{bms} ($n=M=20, L_{\text{max}}=200$ kbits)}
    \addplot[densely dotted,color=purple] coordinates {
            (1000, 130.58) (1100, 137.76) (1200, 143.52) (1300, 150.64) (1400, 157.77) (1500, 162.88) (1600, 168.11) (1700, 173.52) (1800, 178.51) (1900, 182.25) (2000, 186.69)
    };\addlegendentry{\textsc{bms} ($n=10, M=20, L_{\text{max}}=200$ kbits)}
    \addplot[loosely dashed,color=olive,mark=triangle,mark options={scale=1,solid}] coordinates {
            (1000, 134.5) (1100, 141.18) (1200, 147.75) (1300, 154.74) (1400, 161.49) (1500, 167.08) (1600, 172.69) (1700, 178.62) (1800, 185.23) (1900, 190.02) (2000, 195.36)
    };\addlegendentry{\textsc{bms} ($n=M=40, L_{\text{max}}=200$ kbits)}
    \end{axis}
    \end{tikzpicture}
  }
  \caption{Impact of large $m$ and $M$ for the single frame model.}
  \label{fig:5}
\end{figure}
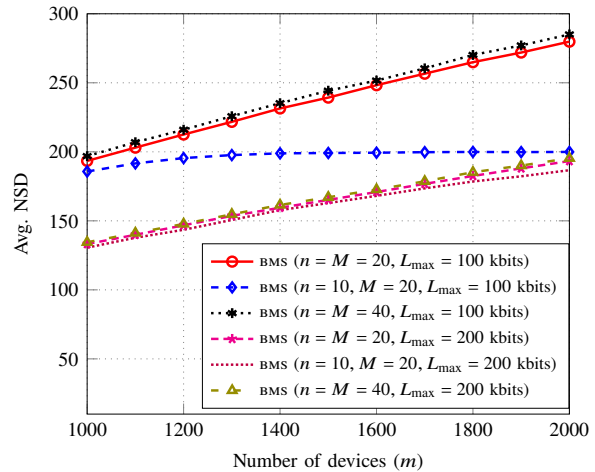%
Fig.~\ref{fig:5} shows the performance of \textsc{bms} for large $m$ and $M$. (Comparison with other algorithms is missing due to running time issues and incompatibility with large $M$.) When the minimum rate requirements $L_{\text{max}}$ is large, increasing $M$ does not help improve the performance even for different $n$ because the interference inside a NOMA group will become large, and thus no more devices can be admitted. However, when the $L_{\text{max}}$ is not very large, then increasing $n$ and $M$ can help improve the performance. When the network is really dense, it is not beneficial to increase $n$ or $M$ very largely. Due to increased SIC complexity and to the limited performance improvements, it is better to keep the values of $n$ and $M$ not very large, e.g., when $L_{\text{max}}=100$ kbits, $n=M=20$ serves about $13.95\%$ of the devices but $n=M=40$ serves about $14.25\%$ of the devices. It is thus better to choose $n=M=20$ rather than $n=M=40$ (the latter gives a gain of only $0.3\%$).

\begin{figure}[ht!]
  \centering
  \captionsetup{justification=centering,margin=1cm}
  \resizebox{.44\textwidth}{!}{%
    \begin{tikzpicture}
    \tikzset{every pin/.style={draw=black,fill=yellow!20,rectangle,rounded corners=3pt,font=\scriptsize},}
    \begin{axis}[
    xlabel={Minimum rate requirements ($L_{\text{max}}$ in kbits)},
    ylabel={Avg. NSD},
    set layers,
    grid=both,
    legend cell align=left,
    xmin=20,
    xmax=100,
    ymin=1,
    ymax=80,
    xtick={20,40,...,100},
    x label style={font=\footnotesize},
    y label style={font=\footnotesize}, 
    ticklabel style={font=\footnotesize},
    legend style={at={(.99,.85)},anchor=east,font=\scriptsize},
    ]
    \addplot[dashed,color=black,mark=asterisk,mark options={scale=1,solid}] coordinates {
            (20, 55.59) (40, 54.94) (60, 54.08) (80, 52.34) (100, 49.21)
    };\addlegendentry{\textsc{opt}}
    \addplot[dashed,color=blue,mark=o,mark options={scale=1,solid}] coordinates {
            (20, 53.4) (40, 51.45) (60, 48.8) (80, 44.44) (100, 40.09)
    };\addlegendentry{\textsc{bms}}
    \addplot[dashed,color=magenta,mark=halfsquare right*,mark options={scale=1,solid}] coordinates {
            (20, 40.17) (40, 38.37) (60, 36.03) (80, 33.74) (100, 31.95)
    };\addlegendentry{\textsc{ath}}
    \addplot[dashed,color=red,mark=triangle,mark options={scale=1,solid}] coordinates {
            (20, 54.31) (40, 52.5) (60, 49.22) (80, 44.28) (100, 40.86)
    };\addlegendentry{\textsc{zz}}
    \addplot[dashed,color=black,mark=asterisk,mark options={scale=1,solid}] coordinates {
            (20, 19.89) (40, 19.69) (60, 20) (80, 19.99) (100, 19.98)
    };
    \addplot[dashed,color=blue,mark=o,mark options={scale=1,solid}] coordinates {
            (20, 19.99) (40, 19.99) (60, 19.98) (80, 19.96) (100, 19.93)
    };
    \addplot[dashed,color=magenta,mark=halfsquare right*,mark options={scale=1,solid}] coordinates {
            (20, 18.1) (40, 18.09) (60, 18.08) (80, 18.0) (100, 17.82)
    };
    \addplot[dashed,color=red,mark=triangle,mark options={scale=1,solid}] coordinates {
            (20, 19.4) (40, 19.35) (60, 19.32) (80, 19.26) (100, 19.18)
    };
    \node[coordinate] (A) at (axis cs:80,42.6) {}; 
    \node[coordinate,pin={[align=left,pin distance=0.8cm,pin edge={black,thick}]215:{$n=30$}}] at (axis cs:77.8,39.95) {}; 
    \node[coordinate] (B) at (axis cs:40,19.5) {}; 
    \node[coordinate,pin={[align=left,pin distance=0.8cm,pin edge={black,thick}]330:{$n=10$}}] at (axis cs:42.2,19.13) {}; 
    \end{axis}
    \draw[black,thick] (A) ellipse (0.2 and 0.8); 
    \draw[black,thick] (B) ellipse (0.2 and 0.2); 
    \end{tikzpicture}
  }
  \caption{Impact of $L_{\text{max}}$ for the single frame model.}
  \label{fig:4}
\end{figure}
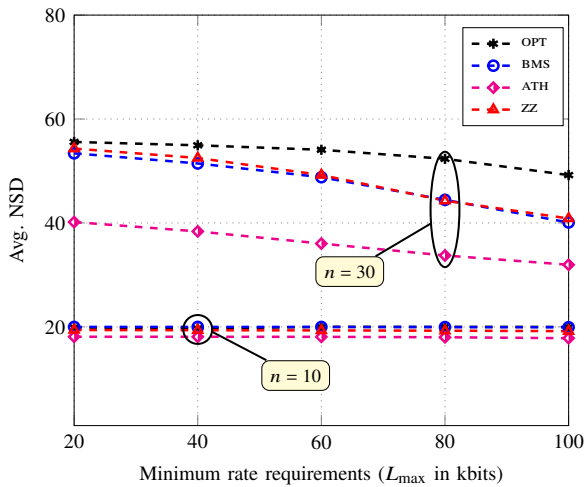
Fig.~\ref{fig:4} illustrates the impact of the minimum rate requirements $L_{\text{max}}$ on the performance of the algorithms for $m=60$. When the number of RBs $n$ is small, the NSD slightly decreases for large values of $L_{\text{max}}$. However, when $n$ is large, the NSD decreases faster with $L_{\text{max}}$. This is because when $n$ is large and $L_{\text{max}}$ is small, an important number of devices can be grouped using NOMA (almost $80\%$ of the devices are served). As soon as $L_{\text{max}}$ increases, some important number of devices will be unsatisfied, and thus the performance drops. Nonetheless, the rate of dropping is much better when $n$ is small since very few devices are grouped using NOMA (almost $30\%$ of the devices are served) for small $L_{\text{max}}$. Thus, unless $L_{\text{max}}$ is not large, NOMA can help serve the same \textit{few} number of devices with different rate requirements. On the other hand, NOMA can serve a larger number of devices but is more influenced by their stringent rate requirements. This remark was also derived in~\cite{8764608} when comparing NOMA and OMA.

\begin{figure}[ht!]
  \centering
  \captionsetup{justification=centering,margin=1cm}
  \resizebox{.44\textwidth}{!}{%
    \begin{tikzpicture}
    \tikzset{every pin/.style={draw=black,fill=yellow!20,rectangle,rounded corners=3pt,font=\scriptsize},}
    \begin{axis}[
    xlabel={Number of devices ($m$) per frame},
    ylabel={Avg. NSD},
    set layers,
    grid=both,
    legend cell align=left,
    xmin=50,
    xmax=300,
    ymin=50,
    ymax=400,
    xtick={50,100,...,300},
    x label style={font=\footnotesize},
    y label style={font=\footnotesize}, 
    ticklabel style={font=\footnotesize},
    legend style={at={(.22,.88)},anchor=east,font=\scriptsize},
    ]
    \addplot[color=blue,mark=o,mark options={scale=1,solid}] coordinates {
        (50, 98.08) (100, 190.02) (150, 262.24) (200, 317.4) (250, 354.5) (300, 376.0)
    };\addlegendentry{\textsc{pl}}
    \addplot[dashed,color=magenta,mark=asterisk,mark options={scale=1,solid}] coordinates {
        (50, 86.04) (100, 155.86) (150, 212.62) (200, 248.38) (250, 274.94) (300, 297.32)
    };\addlegendentry{\textsc{ql}}
        \addplot[dotted,color=red,mark=triangle,mark options={scale=1,solid}] coordinates {
            (50, 65.02) (100, 86.72) (150, 101.26) (200, 110.46) (250, 115.0) (300, 120.22)
        };\addlegendentry{\textsc{rl}}
    \end{axis}
    \end{tikzpicture}
  }
  \caption{Impact of $m$ for the multi-frame model.}
  \label{fig:6}
  \end{figure}
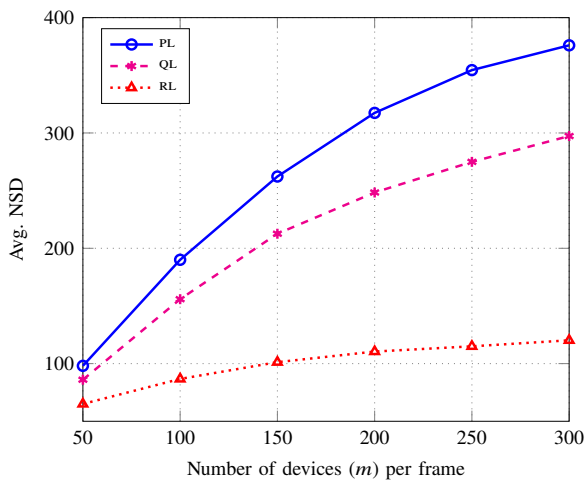

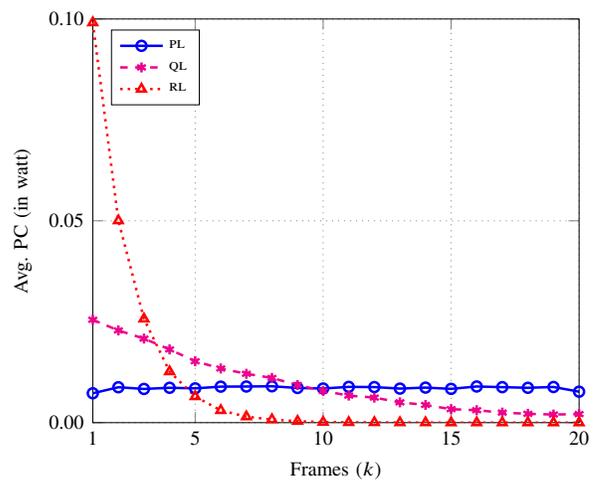
\begin{figure}[ht!]
\centering
  \captionsetup{justification=centering,margin=.7cm}
  \resizebox{.44\textwidth}{!}{%
    \begin{tikzpicture}
    \tikzset{every pin/.style={draw=black,fill=yellow!20,rectangle,rounded corners=3pt,font=\scriptsize},}
    \begin{axis}[
    xlabel={Frames ($k$)},
    ylabel={Avg. PC (in watt)},
    set layers,
    xticklabels={1, 5, 10, 15, 20},
    yticklabels={0.00,0.05,0.10},
    grid=both,
    legend cell align=left,
    xmin=1,
    xmax=20,
    ymin=0,
    ymax=0.1,
    ytick={0,0.05,0.1,0.15,0.2},
    xtick={1,5,10,15,20},
    x label style={font=\footnotesize},
    y label style={font=\footnotesize}, 
    ticklabel style={font=\footnotesize},
    legend style={at={(.22,.88)},anchor=east,font=\scriptsize},
    ]
    \addplot[color=blue,mark=o,mark options={scale=1,solid}] coordinates {
        (1, 0.007280000000000008) (2, 0.008773333333333333) (3, 0.008351111111111103) (4, 0.008640000000000002) (5, 0.008497777777777788) (6, 0.008924444444444452) (7, 0.00895555555555556) (8, 0.00901333333333334) (9, 0.008595555555555552) (10, 0.008462222222222234) (11, 0.008862222222222228) (12, 0.008813333333333341) (13, 0.008440000000000008) (14, 0.008684444444444455) (15, 0.008373333333333342) (16, 0.00896000000000001) (17, 0.008795555555555562) (18, 0.008635555555555554) (19, 0.008835555555555558) (20, 0.007675555555555561)
    };\addlegendentry{\textsc{pl}}
    \addplot[dashed,color=magenta,mark=asterisk,mark options={scale=1,solid}] coordinates {
        (1, 0.025484444444444445) (2, 0.022848888888888887) (3, 0.020826666666666663) (4, 0.018146666666666665) (5, 0.015213333333333334) (6, 0.01342222222222222) (7, 0.012106666666666667) (8, 0.01108) (9, 0.009311111111111112) (10, 0.007862222222222224) (11, 0.006702222222222221) (12, 0.006208888888888889) (13, 0.00504) (14, 0.004386666666666666) (15, 0.003342222222222222) (16, 0.0030844444444444443) (17, 0.0024933333333333335) (18, 0.002195555555555555) (19, 0.002008888888888889) (20, 0.0021199999999999995)
    };\addlegendentry{\textsc{ql}}
        \addplot[dotted,color=red,mark=triangle,mark options={scale=1,solid}] coordinates {
            (1, 0.09910222222222226) (2, 0.050057777777777784) (3, 0.025715555555555554) (4, 0.012657777777777778) (5, 0.00648) (6, 0.00304) (7, 0.0014844444444444445) (8, 0.0007688888888888889) (9, 0.0003733333333333333) (10, 0.0001333333333333333) (11, 7.11111111111111e-05) (12, 5.333333333333333e-05) (13, 2.6666666666666667e-05) (14, 2.222222222222222e-05) (15, 0.0) (16, 8.888888888888888e-06) (17, 4.444444444444444e-06) (18, 0.0) (19, 0.0) (20, 0.0)
        };\addlegendentry{\textsc{rl}}
    \end{axis}
    \end{tikzpicture}
  }
  \caption{Power consumption for the multi-frame model.}
  \label{fig:7}
\end{figure}

In Fig.~\ref{fig:6} and Fig.~\ref{fig:7}, we perform the simulations to learn the power allocation for the multi-frame model (when there are multiple frames). The number of rounds (or episodes) is denoted by $T$. Unless stated otherwise, $k=20$, $n=10$, $m=100$, and $T=100$. The learning rate of \textsc{ql} is $\alpha=0.5$, the \textsc{pl}'s parameters are $\gamma=0.5$, $\beta=0.01$, and $\eta = \gamma/(2(k+1)\sigma_i)$.

  In Fig.~\ref{fig:6}, we plot the total NSD across all frames of different learning algorithms. We remind that, in every frame, there are at most $m$ devices that have packets to send. We compare \textsc{pl} to \textsc{ql} and a very simple learning algorithm, called random learning (\textsc{rl}), which assigns a random amount of power in each frame (from the amount of power left). Learning the transmission power using \textsc{pl} achieves the best performance whereas the worst performance is achieved, as expected, by \textsc{rl} that makes the energy deplete quickly as the number of frames increases due to its random choices. Comparing \textsc{pl} and \textsc{ql}, the performance of the latter degrades as the number of devices increases. The performance gap between \textsc{pl} and \textsc{ql} is about $1.13$ for $m=50$ whereas it becomes about $1.26$ for $m=300$. This is due to the design principle of \textsc{pl} which exploits the problem structure through exploring the TG edges and exploiting the best edges. 
  
In Fig.~\ref{fig:7}, we plot the average power consumption (PC) of all devices across the frames. The PC is averaged over all devices and random realizations and it measures the average power allocation of all devices in each frame. We see that \textsc{rl} depletes its transmission power in the first few frames to end up without energy at later times and thus serves few devices. This is because as the number of frames increases, the random choices available to \textsc{rl} decreases since the sampled power set $[\upper{e}_i]_{\tau_i}$ shrinks. The average PC of \textsc{ql} is much better than \textsc{rl} but still the former allocates more transmission power to the first frames. However, \textsc{pl} allocates the transmission power good enough which gives a good learning outcome. Indeed, \textsc{pl} almost has a uniform average PC across the frames and thus it saves more energy for future frames. Consequently, the power allocation of \textsc{pl} improves the performance by serving the largest number of devices compared to \textsc{ql} and \textsc{rl}.

We conclude that the proposed algorithms \textsc{pl} and \textsc{bms} perform well against the benchmark algorithms for several parameters. Particularly, its performance is very close-to-optimal and is better than the ones of \textsc{ath} and \textsc{zz} when $n$ and $M$ are not very large. For large values of $M$, it is still better than \textsc{ath} and \textsc{zz} but the gap between it and \textsc{opt} increases. Nonetheless, it is still much better than the proven 50\% theoretical performance gap. \textsc{bms} is also close-to-optimal even for massive number of devices. Lastly, \textsc{zz} outperforms \textsc{bms} for large values of $L_{\text{max}}$ or $n$. This is due mainly to the offline nature of the former algorithm. Finally, we can see that the path learning algorithm \textsc{pl} has optimal performances compared to the benchmark learning algorithms \textsc{ql} and \textsc{rl} and it allocates the transmission power evenly across the frames to be able to maximize the NSD in the long run.

\section{Acknowledgement}\label{sec:ack}
We acknowledge the support of the Fonds de recherche du Québec - Nature et technologies (FRQNT) and the Natural Sciences and Engineering Research Council of Canada (NSERC).

\section{Conclusion}\label{sec:conclusions}
In this paper, we studied the online grouping, scheduling, and power allocation problem in beyond 5G cellular IoT NOMA networks. The IoT devices have stringent real-time, rate, and energy requirements. First, we formulated the problem as an integer program. Then, we studied its NP-hardness in different and important cases. To solve the problem efficiently, we proposed online competitive algorithms by first divided it into subproblems of NOMA grouping and scheduling. To obtain the power allocation solution, we used machine learning techniques. Specifically, using Markov decision processes, we modelled the power allocation problem as an online stochastic shortest path problem in directed graphs. Next, we proposed an efficient reinforcement learning algorithm based on exponential-weighting for exploration and exploitation. We showed that our proposed solutions provide close-to-optimal performance in terms of (i) power allocation, and (ii) maximizing the number of IoT devices in a massive access scenario in IoT networks.

\bibliographystyle{IEEEtran}
\bibliography{IEEEabrv,references}

\end{document}